\documentclass[twocolumn,showpacs,preprintnumbers,amsmath,amssymb]{revtex4-1}

\usepackage{graphicx}
\usepackage{dcolumn}
\usepackage{bm}
\usepackage{color}
\usepackage{soul}

\begin{document}


\title{Nonlinear spectral analysis of Peregrine solitons observed in optics and in hydrodynamic experiments}

\author{St\'ephane Randoux}
 \email{stephane.randoux@univ-lille1.fr}
\affiliation{$^1$ Univ. Lille, CNRS, UMR 8523 - PhLAM -
 Physique des Lasers Atomes et Mol\'ecules, F-59000 Lille, France}
\author{Pierre Suret}
\affiliation{$^1$ Univ. Lille, CNRS, UMR 8523 - PhLAM -
  Physique des Lasers Atomes et Mol\'ecules, F-59000 Lille, France}
\author{Amin Chabchoub}
\affiliation{School of Civil Engineering, The University of Sydney, Sydney, NSW 2006, Australia}
\author{Bertrand Kibler}
\affiliation{Laboratoire Interdisciplinaire Carnot de Bourgogne,
  UMR 6303 CNRS-UBFC, Dijon, France}
\author{Gennady El}
\affiliation{$^2$ Department of Mathematical Sciences, Loughborough
 University, Loughborough LE11 3TU, United Kingdom}


\date{\today}

\begin{abstract}
The data recorded in optical fiber \cite{Kibler:10}
and in hydrodynamic \cite{Chabchoub:11} experiments
reported the pioneering observation of nonlinear waves with
spatiotemporal localization similar to the Peregrine soliton
are examined by using nonlinear spectral analysis. 
Our approach is based on the integrable nature of the
one-dimensional focusing nonlinear Schr\"odinger equation (1D-NLSE)
that governs at leading order the propagation of the optical
and hydrodynamic waves in the two experiments. 
Nonlinear spectral analysis provides certain spectral portraits
of the analyzed structures that
are composed of bands lying in the complex plane. The spectral portraits
can be interpreted within the framework of the so-called
finite gap theory (or periodic inverse scattering transform). 
In particular, the number $N$ of bands composing the
nonlinear spectrum determines the genus $g=N-1$ of the solution
that can be viewed as a measure of complexity of the space-time
evolution of the considered solution. Within this setting the ideal,
rational Peregrine soliton represents a special, degenerate genus $2$
solution. While the fitting procedures employed in \cite{Kibler:10}
and \cite{Chabchoub:11}
show that the experimentally observed structures are quite well
approximated by the Peregrine solitons, nonlinear spectral
analysis of the breathers observed both in the optical fiber
and in the water tank experiments reveals 
that they exhibit spectral portraits associated with more general,
genus $4$ finite-gap NLSE solutions. Moreover, the nonlinear spectral analysis
shows that the nonlinear spectrum of the breathers 
observed in the experiments slowly changes with the propagation distance, thus confirming
the influence of unavoidable perturbative higher
order effects or dissipation in the experiments.
  
\end{abstract}

\maketitle

\section{Introduction}

Nonlinear integrable partial differential equations (PDEs) represent an
important class of wave equations that are relevant to many 
fields of physics and applied mathematics
\cite{yang2010nonlinear,akhmediev1997solitons,ablowitz2011nonlinear}.
Notable examples include the
one-dimensional nonlinear Schr\"odinger equation (1D-NLSE),
the Korteweg de Vries (KdV) equation and
the Benjamin-Ono equation. These integrable PDEs
can be solved by using the Inverse Scattering Transform (IST) method
\cite{Zakharov:72,Ablowitz:74},  and they exhibit soliton solutions,
the most celebrated one
being  the propagation-invariant hyperbolic secant 
soliton first discovered by Zabusky and Kruskal through numerical simulations
of the KdV equation \cite{Zabusky:65,Trillo:16}. Solitons represent
solutions of fundamental importance that have the remarkable
property to interact elastically and to preserve (asymptotically)
their shape and velocity upon nonlinear interactions with other
solitons \cite{yang2010nonlinear}.  

Besides fundamental solitons that live on a zero intensity background,
the 1D-NLSE with a self-focusing nonlinearity is known to exhibit
a special class solutions named breathers or solitons on finite 
background (SFB). Some prototypical SFB solutions of the focusing 1D-NLSE
like the Peregrine soliton (PS), the Kuznetsov-Ma (KM) soliton or
the Akhmediev breather (AB) were found  around 
the 80's \cite{kuznetsov1977solitons,ma1979perturbed,peregrine1983water,akhmediev1985generation}. These specific SFB have been experimentally observed 
in a series of optics and hydrodynamic experiments that have been 
realized about thirty years later, 
around 2010 \cite{Kibler:10,Kibler:12,Hammani:11,Chabchoub:11,chabchoub:14}. 
The localization properties of those SFB have recently attracted significant
interest in the context of studies related to the 
formation of rogue waves (RWs), a topic of great interest in
current experimental and theoretical research \cite{Onorato:13}. In particular,
the PS exhibits properties of localization both in space and time that
make it a particularly attractive model of RWs \cite{Shrira:10}.
Even though the PS represents a solution of the 1D-NLSE
that may emerge from the process of the development of modulation instability
of a plane wave, recent works have demonstrated that the PS
also represents a universal nonlinear coherent structure
that emerges from the local regularization of a gradient
catastrophe \cite{Dubrovin:09,Bertola:13,Tikan:17}.
Note that in addition to the PS, there is an infinite
hierarchy of higher-order breather solutions of the 1D-NLSE
that are localized both in space and time while having 
a high peak amplitude \cite{Akhmediev:09,Akhmediev:09c}.
These higher-order SFB solutions have
been observed in some recent optics and hydrodynamic experiments
\cite{Frisquet:13,Chabchoub:12,Chabchoub:12b}. Also,
some ``superregular solitonic solutions'' which appear as small
localized perturbations of a plane wave at a certain moment of time have been shown to
describe a possible scenario of the nonlinear stage of the modulation instability of the
plane wave \cite{Zakharov:13,Kibler:15}. Another scenario
of the development of modulational instability induced by a localized
perturbation theoretically described in \cite{Biondini:16a} was
recently observed in a fiber optic experiment \cite{Kraych:18}.

In most of the experiments that have reported so far the observation of SFBs,
the signal to noise ratio is so good 
that it is fully appropriate to compare the traces  recorded experimentally
with the analytical expression describing the SFB 
that has been measured. As a result, in most of the hydrodynamic and optics
experiments, the discrepancy between the analytical expression and the
experimental data is so small that it is naturally
concluded that features typifying the SFB under interest have
been indeed experimentally observed \cite{Kibler:10,Kibler:12,Chabchoub:11,chabchoub:14,Frisquet:13,Chabchoub:12,Chabchoub:12b,Kibler:15}.
Although the situation is less straigtforward, similar
studies have also been realized in experiments
or numerical simulations where the breather-like structures may
emerge randomly in space and time from some random initial condition.
The question of the identification of these breather-like structures 
represents an issue of importance in the field of integrable 
turbulence \cite{Zakharov:09,Randoux:14,Walczak:15,Agafontsev:15,Randoux:16,Randoux:17,Soto:16,Akhmediev:16}. In this context, fitting procedures
have been extensively employed in attempts to identify breather structures
that emerge from a randomly fluctuating background
\cite{Dudley:14,Akhmediev:09c,Akhmediev:09b,Toenger:15,Walczak:15,Suret:16,Nahri:16,Tikan:18}. 

In this paper, we use nonlinear spectral analysis as a mathematical tool
to examine the nature of the PS-identified events that have been
observed in optics and hydrodynamic experiments reported in 
ref. \cite{Kibler:10} and \cite{Chabchoub:11}.
Our approach, which is based 
on the integrable nature of the 1D-NLSE, consists in computing 
a {\it spectral} (IST) portrait of the experimentally observed
coherent structure that is considered as a local 
solution of the propagation equation. Even though the coherent structures 
experimentally observed in ref. \cite{Kibler:10} and \cite{Chabchoub:11} 
are reasonably well fitted by the rational mathematical expression
defining the PS, they are found to display spectral signatures
that depart from the spectral signature typifying the 
pure PS defined by its rational mathematical expression. 
The noticeable distortions that are evidenced by the spectral IST
analysis arise from the small discrepancies 
existing between experimental signals and the ideal mathematical expression 
defining the PS (see Eq. (\ref{PS})). Even though the PSs
experimentally observed retain
some degree of proximity with the pure PS, the spectral analysis
reveals that they represent more complex solutions of the 1D-NLSE
that can be expressed using Riemann theta functions in the
framework of the finite-gap theory \cite{Osborne2010nonlinear}. 

This paper is organized as follows. Sec. \ref{ist_theory} introduces
the theoretical framework and describes the mathematical tools
of nonlinear spectral analysis that are based on the inverse
scattering method. Sec. \ref{numerical_ist} presents the numerical
methods that are used to practically implement nonlinear spectral
analysis of SFBs. Sec. \ref{sec:PS_opt} and
Sec. \ref{sec:PS_hydro} present the results obtained from
the nonlinear spectral analysis of PS-like coherent structures
observed in optics
\cite{Kibler:10} and hydrodynamic \cite{Chabchoub:11} experiments,
respectively. A brief summary and a conclusion of our work are
given in Sec. \ref{conclusion}.

\section{The inverse scattering transform method for the nonlinear 
  spectral analysis of the Peregrine soliton} \label{sec:IST}

\subsection{Theoretical framework}\label{ist_theory}

We consider the integrable focusing 1D-NLSE in the following form: 
\begin{equation}\label{nlse}
  i \psi_t +  \psi_{xx} +  2 \,  |\psi|^2 \, \psi=0,
\end{equation}
where $\psi(x,t)$ represents the complex envelope of the 
optical or hydrodynamical wave fields.

In the experiments considered in Sec. \ref{sec:PS_opt} and 
in Sec. \ref{sec:PS_hydro}, 
the evolution variable is not the time $t$ but the 
longitudinal coordinate $z$ measuring 
the propagation distance along the optical fiber or along the 
$1D-$flume. At a given position inside the fiber or inside the 
flume, the field 
does not change in space $x$ but in time $t$. Therefore the 1D-NLSE
governing wave propagation inside the fiber or inside the 1D-flume is obtained 
by performing the following changes of variables: $t \leftrightarrow z$, 
$x \leftrightarrow t$ together with appropriate scaling of 
the NLSE coefficients with parameters
typifying either optics or hydrodynamic
experiments \cite{Chabchoub:15,Chabchoub:16}, see Appendix \ref{appendixa} and
\ref{appendixb}.

In the IST method, the focusing 1D-NLSE is represented as a
compatibility condition of two linear equations \cite{Zakharov:72},
\begin{equation}\label{LP1}
Y_x=
 \begin{pmatrix}
-i \xi & \psi \\ -\psi^* & i \xi \\
\end{pmatrix} 
Y=Q(x,t,\xi) \, Y, 
\end{equation}
\begin{equation}\label{LP2}
Y_t=
 \begin{pmatrix}
-2 i \xi^2+i|\psi|^2 & i \psi_x +2\xi \psi \\ i \psi_x^* - 2\xi \psi^*
& 2 i \xi^2 -i|\psi|^2 \\
\end{pmatrix} 
Y,
\end{equation}
where $\xi$ is a (time-independent) complex spectral parameter and $Y(x,t,\xi)$ is a
vector. The spatial linear operator (\ref{LP1}) and the temporal
linear operator (\ref{LP2}) form the Lax pair of Eq. (\ref{nlse}).
For a given potential $\psi(x,t)$ the problem of finding the spectrum
$\sigma[\psi]$ and the corresponding scattering solution $Y$ specified
by the spatial equation (\ref{LP1}) is called 
the Zakharov-Shabat (ZS) scattering problem \cite{yang2010nonlinear}. 
The ZS problem is formally analogous to
calculating the Fourier coefficients in Fourier theory
of linear systems.

Reformulating Eq. (\ref{LP1}) as 
\begin{equation}\label{LP3}
 \begin{pmatrix}
\partial_x +i \xi & -\psi \\ \psi^* & \partial_x - i \xi \\
\end{pmatrix} 
Y= {\mathcal L}^{(x)} Y =0, 
\end{equation}
every NLSE solution $\psi$, which rapidly decay as $|x| \to \infty$,
is characterized by the spectrum 
$\sigma[\psi]=\{\xi \in \mathbb{C} | \mathcal{L}^{(x)} Y =0, |Y| \, \text{bounded} \, \forall x \}$ of the
linear operator $\mathcal{L}^{(x)}$. 
The discrete eigenvalues
of the ZS operator $\mathcal{L}^{(x)}$ with decaying potentials give spectral portraits that provide
precise IST signatures of fundamental soliton solutions of
Eq. (\ref{nlse}) that exponentially decay to zero
as $|x| \to \infty$ \cite{yang2010nonlinear},
while the continuous spectrum component of $\psi$ lies on the real axis.

SFBs are not solutions of Eq. (\ref{nlse}) that decay to zero
as $|x| \to \infty$ but they rather generically represent certain limits of periodic or quasi-periodic 
solutions of Eq. (\ref{nlse}) obtained in the framework of the extension of IST called the finite-gap theory \cite{Tracy:84, Osborne2010nonlinear,Grinevich:17}. 
For periodic boundary conditions, i.e. $\psi(x+L,t)=\psi(x,t) \, \forall t $,
the spectrum has to be determined from
Floquet spectral theory \cite{Easthambook}. 
In this framework, the spectrum of $\psi$ is expressed
in terms of the transfer matrix
$M(x+L;\psi,\xi)$ across a period, where $M(x;\psi,\xi)$
is a fundamental solution matrix of the Lax pair.
Introducing the Floquet discriminant
$\Delta(\psi,\xi)=Tr[M(x+L;\psi,\xi)]$ as the trace
of the transfer matrix, the (Floquet) spectrum of a 
solution of Eq. (\ref{nlse}) fulfilling periodic boundary conditions
is given by \cite{Ablowitz:01,Islas:05,Calini:12}
\begin{equation}\label{spectrum}
\sigma[\psi]=\{\xi \in \mathbb{C} |
\Delta(\psi,\xi) \in \mathbb{R}, -2 \leqslant \Delta(\psi,\xi)
\leqslant 2\}
\end{equation}
The Floquet spectrum of a periodic solution of Eq. (\ref{nlse}) 
typically consists of bands lying in the complex plane. 
The number $N$ of these bands
determines the  genus $g=N-1$ of the finite-gap solution. The genus $g$ solution can be 
represented in the form: 
\begin{equation}\label{theta_solution}
  \psi_g(x,t)=q \frac{\Theta_g(x,t,{\bf \nu_{-}^{0})}}{\Theta_g(x,t,{\bf \nu_{+}^{0}})} e^{i q^2 t}
\end{equation}
where $q \in \mathbb{R}$ and $\Theta_g$ is the Riemann theta
function of genus $g$ \cite{Osborne2010nonlinear,El:16}.
${\bf \nu_{+,-}^{0}} \in \mathbb{R}^g$ are called the phases, which
are defined by the initial condition.

\begin{figure}[h]
\includegraphics[width=9cm]{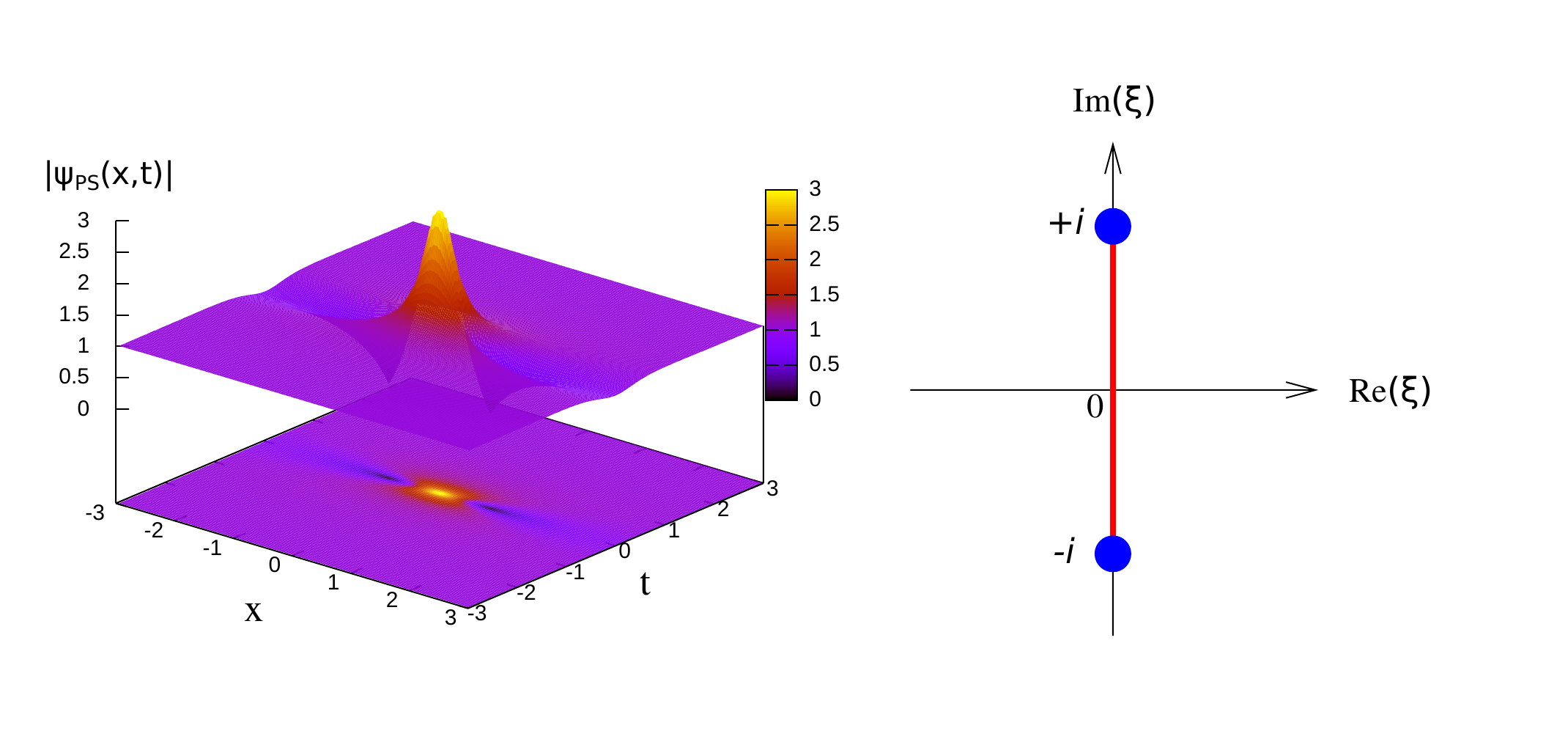}
\caption{Spatiotemporal evolution of the PS (left) and corresponding 
time-independent spectral portrait (right) plotted in the complex 
plane and obtained from finite-gap theory (periodic IST). 
The red line in the spectral portrait represents the so-called branchcut, or the spine,  and the 
blue points represent complex conjugate double points of the spectrum.}
\label{fig:ps_fgt}
\end{figure}

From the perspective of periodic IST or finite-gap theory, 
the plane wave  $\psi =qe^{2iq^2 t}$ 
is a genus $0$ solution of the 1D-NLSE. It has the spectrum
represented by one band or a `branchcut' (also called a `spine' in \cite{Tracy:84, Osborne2010nonlinear}) lying in the complex plane
between two points $iq$ and $-iq$ of the simple spectrum of
the {\it periodic} ZS problem \cite{ma1981periodic,Tracy:84}.
The solutions of genus $1$ of the 1D-NLSE with periodic boundary conditions 
are cnoidal waves which have their spectrum composed 
of $2$ bands lying in the complex plane \cite{El:16}.  
The fundamental bright soliton $\psi_S(x,t)=\hbox{sech}(x,t) e^{2it}$ 
represents a limit solution of genus $1$ obtained when the 
period of the cnoidal wave tends to infinity. 
In this limiting case, the size of the spectral bands tends to zero 
and the spectrum of the stable bright fundamental soliton 
$\psi_S$ living on a zero-background is simply made of two 
doubly-degenerate complex conjugate eigenvalues 
$\xi_{\pm}=​​\pm i/2$ \cite{Randoux:16b}. Note that the eigenvalues of 
a fundamental soliton moving with some nonzero velocity $v$ in 
the $(x,t)$ plane become $\xi_{\pm}=​​-v/4 \pm i/2$ \cite{yang2010nonlinear}.

The standard SFBs (ABs, KM solitons and PSs \cite{Dudley:14}) all represent degenerate genus $2$
solutions of the NLSE with periodic boundary conditions.
This means that their IST spectra are composed of three bands, two of which undergo degeneracy, giving rise to two double points (see
ref. \cite{Randoux:16b} for a detailed description of the spectra
of the standard SFBs). There are two ways  this degeneracy can occur. 
The first one is realised when the three bands composing the spectrum of a genus-2 solution all sit  on the vertical imaginary axis in the complex plane.  In particular, the (time-independent)         
IST spectrum of a PS that lives on a unity background can be viewed as being made of
 three identical bands lying along the vertical imaginary axis
between the points $+i$ and $-i$ of the complex plane so that two  bands ``annihilate'' each other giving rise to two double points (analogs of soliton spectra) that are located exactly at the endpoints $\pm i$ of  the remaining third band
crossing the real axis and corresponding to the plane wave background (see Fig. 1). Another way of the ``spectral'' construction of the PS  out of a genus 2 solution  is to consider the double points as  two Schwartz-symmetrical  (``rogue mode'') bands collapsed onto the branch points of the ``Stokes mode'' band between $+i$ and $-i$. The way the PS spectrum is formed in reality depends on the specific configuration of initial conditions (see \cite{El:16} for a relevant discussion). The two above interpretations of the PS spectrum  also indicate two possible ways its deformation, under the effects of perturbation, when the degeneracy is removed. These spectrum deformations are particularly relevant to the subject of this paper.

\subsection{Numerical determination of the IST spectrum of the Peregrine
soliton}\label{numerical_ist}

Although the spectral portraits of SFBs given by the IST
theory \cite{Biondini:14,Biondini:15,Gelash:14} are standard, the more general wave
structures are often difficult to analyze, and some numerical
procedures have been developed to compute their IST spectra
\cite{Boffetta:92,yang2010nonlinear,Yousefi:14,Wahls:15,Frumin:15,Kamalian:16,Turitsyn:17}.
The numerical IST methods can be regarded as 
powerful tools for the nonlinear Fourier analysis of waveforms
in situations where the propagation is described by integrable
equations like the KdV equation or the 1D-NLSE. 
Such tools have been successfully implemented many times in 
the context of nonlinear analysis of hydrodynamic random wave trains, see for instance ref.
\cite{Slunyaev:06,Osborne:93,Osborne:95,Osborne:95b,Osborne:14,Osborne2010nonlinear,Ablowitz:01,Islas:05,Calini:12,Islas:11}. 
More recently, the use of nonlinear Fourier transform has been promoted as
a possible way to overcome transmission limitations in
fiber communication channels by encoding information in the nonlinear
IST spectrum \cite{Prilepsky:14,Frumin:17,Turitsyn:17}.
Note that soliton radiation beat analysis represents another numerical
technique that permits to determine the soliton content of
light pulses \cite{Bohm:06,Bohm:07,Mitschke:17}. 

In this work, we use two different algorithms to
perform the numerical IST analysis of PSs that have been experimentally
recorded in two different media as reported in ref. \cite{Kibler:10,Chabchoub:11}.
Both algorithms assume that periodic boundary conditions are
applicable. The first algorithm is
based on a method introduced by Boffetta and Osborne (BO) 
in 1992 \cite{Boffetta:92}.
The second algorithm, named Fourier collocation method \cite{yang2010nonlinear},
requires the implementation of a procedure in which the analyzed wavetrain
is periodized before the IST analysis is made. This periodization step
is unnecessary in BO's method where periodicity
of the analyzed wavetrain is implicitly assumed. In this Section,
we briefly describe the two methods and we show how they do apply
to the PS defined by its rational mathematical expression. 

BO's method relies on the idea that
Eq. (\ref{LP1}) is a first order system that can
be integrated with respect to $x$, thus yielding
\begin{equation}\label{LP1_solution}
  Y(x,t^*,\xi)=Y(x_0,t^*,\xi) \exp \left( \int_{x_0}^{x} Q(x',t^*,\xi) dx' \right),
\end{equation}
$x_0$ being an arbitrary point inside the interval $[0,L]$
where the potential $\psi(x,t^*)$ is defined at a given time $t^*$.
In BO's numerical procedure \cite{Boffetta:92}, the potential
$\psi(x,t^*)$ is approximated by a piecewise-constant potential by using a
conventional discretization procedure. To this end, the box of size $L$
is sampled into an ensemble of $N+1$ points at the positions
$x_j=j \Delta x$ where $j$ is an integer ($j=0,...,N$) and $\Delta x=L/N$. 
Assuming that $\Delta x \ll 1$, we immediately have
\begin{equation}\label{LP1_solution_approx}
  Y(x_{j+1},t^*,\xi)=\exp \left( Q(x_j,t^*,\xi) \, \Delta x  \right) Y(x_{j},t^*,\xi).
\end{equation}
Iterating the latter relation over one spatial period gives
$Y(x_0+L,t^*,\xi)=M(x_0+L,t^*,\xi) \,\,\, Y(x_0,t^*,\xi)$
where
\begin{equation}\label{transfer_matrix}
M(x_0+L,t^*,\xi)=\prod_{j = N}^{0} U(x_j,t^*,\xi).
\end{equation}
with $U(x_j,t^*,\xi)=\exp \left( Q(x_j,t^*,\xi) \, \Delta x  \right)$.
The elements of the matrix $U(x_j,t^*,\xi)$
are explicitly given in ref. \cite{Boffetta:92} in terms of
hyperbolic trigonometric functions. The matrix $M$ given by Eq. (\ref{transfer_matrix})
represents a numerical approximation of the transfer matrix of the Floquet
theory (see Sec. \ref{ist_theory}). Once the transfer matrix
is computed numerically for the piecewise approximation
of the potential $\psi(x,t^*)$, it is straightforward to
determine the Floquet spectrum characterizing
$\psi(x,t^*)$ by computing the trace $\Delta$ of $M$
and applying the definition given in (\ref{spectrum}).

In the so-called Floquet collocation method, the determination of discrete
eigenvalues $\xi$ of the ZS
system is made by rewriting Eq. (\ref{LP1}) as a standard linear
eigenvalue problem 
\begin{equation}\label{eigen}
 \begin{pmatrix}
- \partial_x & \psi(x,t^*) \\ \psi^*(x,t^*) & \partial_x \\
\end{pmatrix} 
Y=i \xi Y.
\end{equation}
The $x-$axis is truncated into a box of finite size $L$. The
eigenvector $Y=(y_1(x,t^*),y_2(x,t^*))^T$ as well as the potential $\psi(x,t^*)$
at a given time $t^*$ are expanded into Fourier series.
These Fourier expansions are substituted in Eq. (\ref{eigen}) and the
obtained system for the eigenvalues is then solved by using standard linear
algebra routines \cite{Kamalian:16,yang2010nonlinear}.
In other words, the Fourier collocation method represents a reformulation
of the ZS problem as an eigenvalue problem in the Fourier space. 
Even though the Fourier collocation method is not appropriate in the context 
of eigenvalue communication \cite{Turitsyn:17}, it has been shown to provide accurate spectral
signatures of all the standard SFB, see Fig. 2 of ref. \cite{Randoux:16b}. 

Now that the numerical IST methods have been presented, we focus on the question
of the numerical IST analysis of the PS described by its well-know rational
mathematical expression \cite{peregrine1983water,Kibler:10,Chabchoub:11}
\begin{equation}\label{PS}
\psi_{PS}(x,t)=\left( 1- \frac{4(1+4it)}{1+4x^2+16t^2} \right) \, e^{2it}.
\end{equation}
First, it should be emphasized that the solution given by
Eq. (\ref{PS}) lives on the infinite line and that any numerical
IST analysis imposes some truncation of the solution (\ref{PS})
that becomes confined inside a box having a finite size $L$. This truncation immediately 
corrupts the IST spectrum that is no longer perfectly composed
of three degenerate bands lying between the points $+i$ and $-i$ in
the complex plane (see the discussion at the end of Sec.~II.A). This effect is illustrated in Fig. 2 which reveals
that the IST spectrum of the truncated PS is made of one main band
crossing the real axis and of two smaller bands crossing the
vertical imaginary axis. The greater  the size $L$
of the numerical box is, the smaller is the size of these two bands. 
In fact, numerical simulations show that the positions of the
endpoints delimiting the three bands converge towards 
$\pm i$ when $L$ increases, as it is expected from the finite-gap theory. 

\begin{figure}[h]
\includegraphics[width=9.2cm]{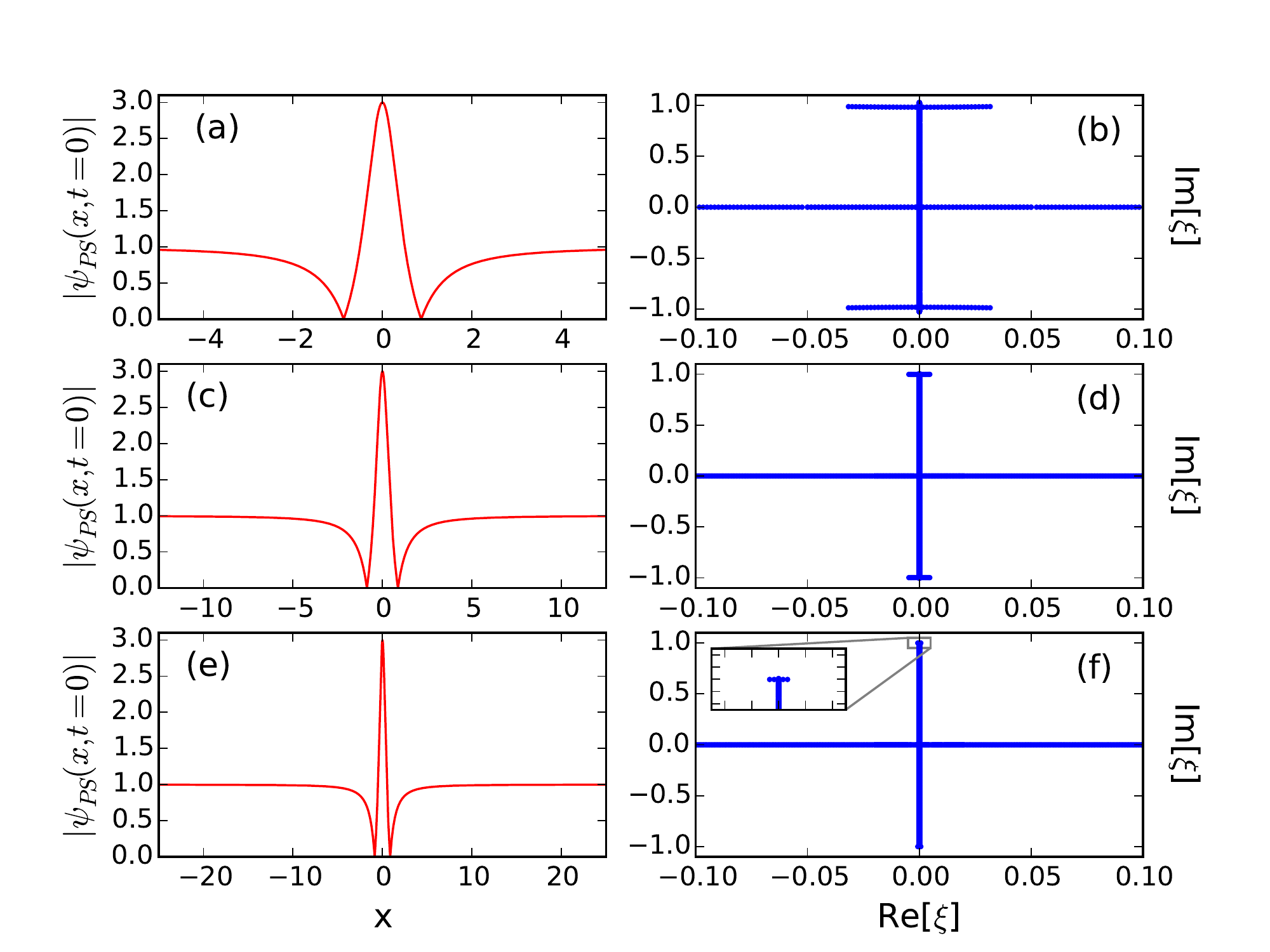}
\caption{Spatial profiles (left) and corresponding spectral portraits 
(right) of the PS given by Eq. (\ref{PS}). 
The spectral portraits have been computed with numerical boxes 
having sizes $L=10$ (b); $L=25$  (d) and $L=50$  (f).}
\label{fig:ist_rational_ps}
\end{figure}

Even though the IST spectrum of the PS is slightly corrupted 
by the truncation procedure, 
the computed spectral portraits preserve the global information 
that the analyzed structure is the PS. First, the computed spectra
are composed of three bands, which means that the analyzed structure 
is a genus $2$ solution of Eq. (\ref{nlse}). Then, the computed IST 
spectra have a symmetry with respect to the vertical imaginary axis that 
is reminiscent of the degenerate nature of the analyzed structure. 
Finally, the endpoints delimiting the three computed bands 
are close to the correct theoretical positions $\pm i$ in 
the complex plane.

If the truncation of the PS can be simply understood as a 
{\it local} procedure dedicaded to isolate its central core part, 
the subsequent IST analysis made by using BO's method 
or the Fourier collocation method must be understood as 
representing the numerical IST analysis of the truncated PS
of size $L$ that has been expanded to produce a 
periodic pattern of period $L$. 
The BO method indeed provides the IST spectra of potentials 
that are periodic in space. In this method, the period of
the analyzed structure is naturally equal to the size $L$ 
of the numerical box
and the period $L$ is inherently encoded into the trace $\Delta$ of 
the transfer matrix. It is therefore {\it implicit} in BO's method 
that the IST analysis of a {\it periodic} structure
of period $L$ is made. On the other hand, in the Fourier 
collocation method extensively used in ref. \cite{Randoux:16b}, 
the periodization of the truncated PS 
must be made in an {\it explicit} way prior to the IST analysis. 
As discussed in detail in ref. \cite{Randoux:16b}, 
it is mandatory to produce a periodic series of 
truncated PS before solving the ZS problem in Fourier space.  
Our computations show that, as long as the analysis reported in
Fig. 2 is concerned, there is no significant quantitative difference 
between numerical results obtained with the Fourier collocation method and 
with the BO method.

Summarizing, the numerical IST analysis described in 
this Section represents a tool which has been used here 
to perform a local finite-band approximation of the PS
defined by Eq. (\ref{PS}). 
This tools captures the fact that the PS truncated 
to its central core part is {\it locally} composed of three 
dominant nonlinear modes that are embedded within some 
symmetric interaction process. 

\section{Nonlinear spectral analysis of the
  Peregrine soliton observed in optical fiber experiments} \label{sec:PS_opt}

In the experiment reported in ref. \cite{Kibler:10}, a
weakly modulated light wave is injected inside an optical fiber. 
Nonlinear propagation inside the fiber leads to the
generation of a breather structure that exhibits
properties of localization in space and time close to the PS. 
The experimental conditions and the physical parameters
characterizing this optical fiber experiment
are known with a very good accuracy. 
Some important insight into the understanding of
the experimental results reported in ref. \cite{Kibler:10}
can first be recalled by means of numerical simulations.
In Sec. \ref{sec:PS_opt_numerical_analysis}, we report 
numerical simulations in which we analyze in detail the optical
fiber experiment reported in ref. \cite{Kibler:10}.
In Sec. \ref{sec:PS_opt_data_analysis}, we perform the 
nonlinear spectral analysis of the optical signal that has been
recorded in ref. \cite{Kibler:10}.

\subsection{The optical fiber experiment analyzed by numerical simulations}\label{sec:PS_opt_numerical_analysis}
 
As underlined in ref. \cite{Kibler:10}, the PS represents an ideal 
asymptotic limit that can never be reached in practice.
In these circumstances, the approach that was taken in ref. \cite{Kibler:10}
has consisted in demonstrating that the spatiotemporal 
localization properties of the PS are however experimentally 
observable. In the optical fiber experiment reported 
in ref. \cite {Kibler:10}, SFBs have been generated by 
weakly modulating the plane wave used as an initial 
condition. Shifting the frequency of the pertubative  Fourier component 
added to the plane wave while appropriately 
adjusting the optical power, it has been 
shown that the breather structures 
emerging at the nonlinear stage of the development of modulational instability 
exhibit properties of localization 
in space and time that are compatible with the ones characterizing the 
PS, see Fig. 2(b) of ref. \cite{Kibler:10}. 
At the point where the strongest localization effects occur 
in the optical fiber experiment ($a=0.42$ in Fig. 5(d) of 
ref. \cite{Kibler:10}), the intensity profile of 
the experimentally observed localized structure  has 
a core part that fits quantitatively very well the 
profile characterizing the PS mathematically defined by Eq. (\ref{PS}).

Before using the numerical implementation of the periodic IST for the analysis of the experiment 
reported in ref. \cite{Kibler:10}, one important comment 
should be made about the suitability of this mathematical 
tool for the analysis of the data recorded in optical fiber experiments
that take periodic signals as initial condition \cite{Kibler:10,Kibler:12,Hammani:11,Frisquet:13,Frisquet:14}. 
The fact that the initial condition is generated by Fourier synthesis 
in a number of optical fiber experiments has the important consequence
that it is breather structures having some periodicity both in space and 
time that are inherently produced in all these experiments.
In these circumstances, periodic IST naturally represents the appropriate 
tool for the nonlinear analysis of the observed coherent structures.

\begin{figure}[h]
\includegraphics[width=9.cm]{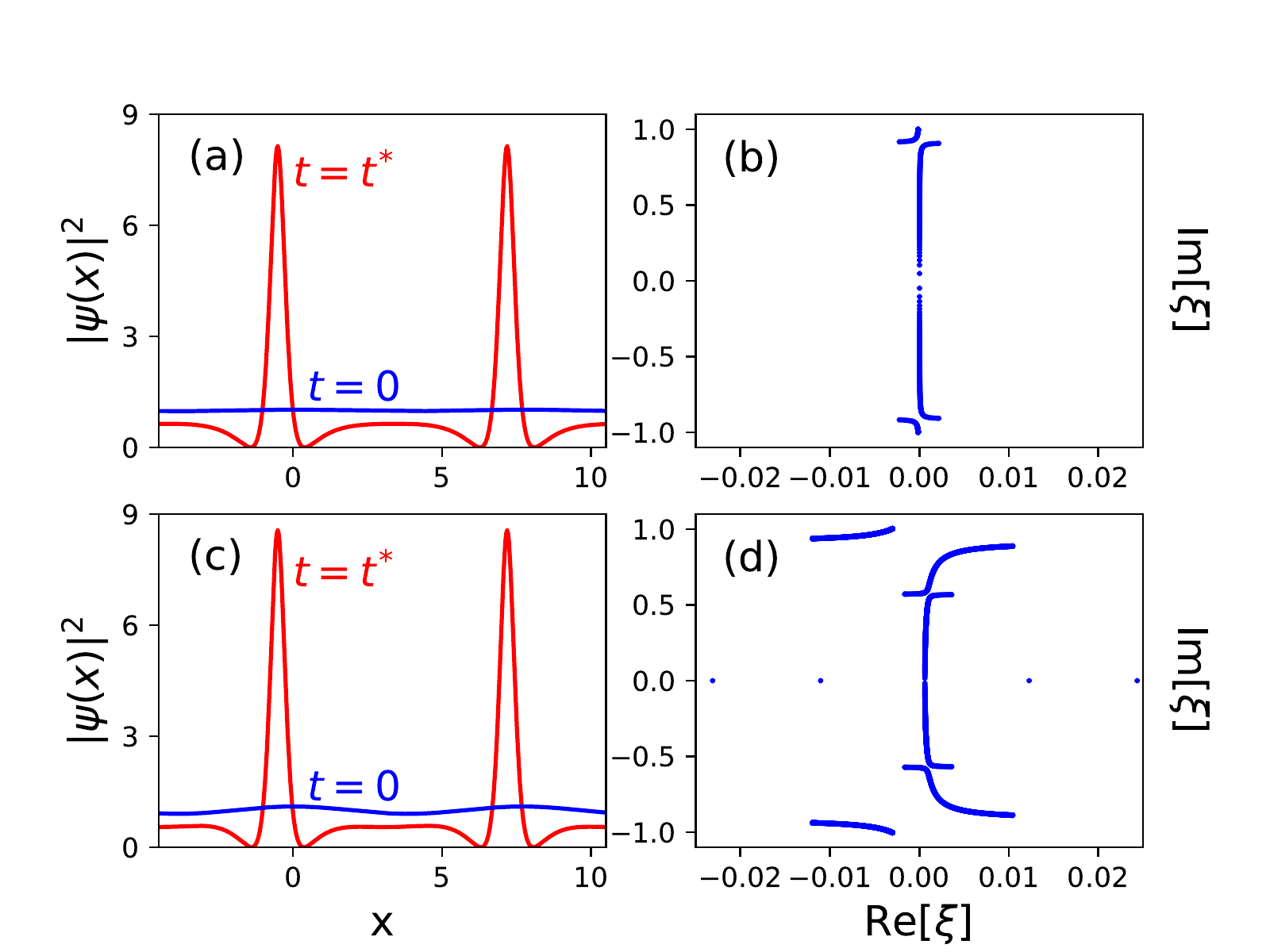}
\caption{Numerical simulations of Eq. (\ref{nlse}) (left column)
  and the corresponding IST spectra (right column) computed 
  with the initial condition given by Eq. (\ref{ci_opt}). Parameters
  are (a) (b) $\epsilon=0.01$, $k_0=0.822$,
  (c), (d) $\epsilon=0.05$, $k_0=0.822$.
  $t^{*}$ represent the times at which the breather structures reach
  their maximum intensities: (a) $t^{*}=3.162$, (c) $t^{*}=2.117$.
  The size $L=7.64$ of the box used for numerical IST analysis
  is equal to the period of the modulated wave ($L=2\pi/k_0$).
}
\label{fig:ist_nlse_opt_exp}
\end{figure}

Taking the 1D-NLSE in the form given by Eq. (\ref{nlse}), 
the field used as initial condition in the experiment of
ref. \cite{Kibler:10} reads 
\begin{equation}\label{ci_opt}
\psi(x,t=0)=\psi_0(x)=1+\epsilon \, \, e^{-i k_0 x}
\end{equation}
with $k_0=2\pi/L$ where $L$ represents the size of the 
box that will be used here for the numerical IST analysis. The 
dimensionless spatial frequency $k_0$ is directly related 
to the frequency detuning between the two lasers 
used in the experiment and it also depends on the nonlinear length and 
on the second-order dispersion coefficient of the fiber,
see Appendix \ref{appendixa}
for details about relations between physical parameters and 
dimensionless parameters. In the experiments reporting the 
so-called maximally compressed pulse (see Fig. 5(d) of 
ref. \cite{Kibler:10}) that is well fitted by Eq. (\ref{PS}),
the numerical values of $k_0$ and $\epsilon$ are $0.822$ 
and $0.225$, respectively. 

Fig. 3(a) shows the intensity profile $\psi(x,t^*)$ of the breather
structure that is obtained from numerical integration of Eq. (\ref{nlse})
by taking the value $k_0=0.822$ typifying the experiment.
On the other hand, the numerical value of the parameter $\epsilon$
is $0.01$, which is much smaller that the experimental value of $0.225$.
With these parameters, the time at which the structure 
reaches its maximum amplitude is $t^{*}=3.162$ which corresponds 
to a virtual propagation distance of $\sim 2.1$ km in an optical fiber
with no losses and with linear and nonlinear parameters identical
to those typifying the fiber used in ref. \cite{Kibler:10}, see Appendix \ref{appendixa}.

Fig. 3(b) shows the IST spectral portrait
of the breather structure shown in Fig. 3(a).
It has been computed by using the Fourier collocation
method but identical numerical results are obtained
by using BO's method. 
Let us emphasize that the spectral portrait is time-independent:
identical IST spectra are obtained by taking the field at $t=0$,
at $t=t^{*}$ or at any other evolution time.
The spectral portrait of Fig. 3(b) being composed of three bands, 
the IST spectral analysis reveals that the generated structure 
represents a genus $2$ solution of the 1D-NLSE
that can be closely related to the AB.

The spatially periodic breather structures shown in Fig. 3(a,b) indeed bear strong resemblance to the ABs, which are typically associated with the development of modulational instability of the plane wave modified by a small-amplitude periodic perturbation \cite{Dudley:09,Hammani:11,Mussot:18}. In this connection let us emphasize that the AB solution represents a pure space-periodic
homoclinic solution of the 1D-NLSE that breathes only once
in time. As illustrated in Fig. 1(a) of ref. \cite{Ablowitz:01},
the IST spectrum of the pure AB is very particular because it is
composed of a branchcut between $+i$ and $-i$ and also of two double
points lying at positions given by $\pm i \, \sqrt{2a}$,  with the conventions
of ref. \cite{Kibler:10}. As discussed in detail in ref. 
\cite{Ablowitz:96,Ablowitz:01}, the modulation of a plane
wave by two weak side bands generally produces more complex solutions
than the pure AB. However, the spectra of the
structures generated by periodic modulation of
the plane wave retain some proximity to the pure AB solution.

Here, by modulating the plane wave with only one side band,
two gaps are opened in the branch cut crossing the real axis, see Fig. 3(b).
The central position of these gaps are $\pm i \, 0.914$
which is very close to the theoretical
value $\pm i \, \sqrt{2 a}=\pm i \, 0.916$
that characterizes the pure AB solution ($a=0.42$) considered in
ref. \cite{Kibler:10}.
The IST spectrum is not symmetric with respect to the vertical
imaginary axis because the Fourier spectrum of the initial
condition is itself not symmetric. This initial asymmetry
induces a slight drift of the structure that moves in the $(x,t)$ plane
with a non-zero velocity.
This drift in the $(x,t)$ plane is readily observable in Fig. 3(a)
where the breather structure
reaches its peak intensity at $(x=-0.51,t^*)$ while the 
initial field intensity is maximal at $(x=0,t=0)$.
Such a breather structure was classified as a ``left'' state in ref. 
\cite{Ablowitz:96,Ablowitz:01}, i. e. a left-travelling
wave solution of the 1D-NLSE.
In ref. \cite{Zakharov:13,Kibler:15,Akhmediev:09}
it was also considered into the class of
the general one-breather solution with an arbitrary group
velocity [26,27,39]. The impact of an asymetric initial excitation
of ABs has been analyzed in Refs.
\cite{Frisquet:13,Akhmediev:09d,shaping_light_book_kibler}.
We note that exact analytical description of
the formation of ABs and related structures in the initial value
problem \eqref{nlse}, \eqref{ci_opt} was also developed in the recent
work \cite{Grinevich:17} using the finite-gap theory.

Increasing the value of $\epsilon$ to $0.05$ while keeping 
the same value $k_0=0.822$ for the wavenumber characterizing the 
small perturbation of the plane wave, 
numerical simulations of Eq. (\ref{nlse}) reveal that the breather 
structure shown in Fig. 3(c) now reaches its
maximum amplitude at $t^{*}=2.117$. Even though the intensity 
profile of the breather structure obtained for $\epsilon=0.05$ 
is similar to the one of the breather structure obtained for $\epsilon=0.01$, 
the spectral portrait reveals that the analyzed structure is no longer 
a genus $2$ solution of the 1D-NLSE but a genus $4$ solution, i.e. the solution involving $g+1=5$ nonlinear ``eigenmodes'' \cite{Tracy:84, Randoux:16}. 
In terms of the IST spectrum, two additional gaps are now open
in the vertical band crossing the 
real axis, which results in a IST spectrum now composed of $5$ bands. 
This change in the genus of the solution implies that some more
complex spatiotemporel evolutions can possibly be observed,
as shown e.g. in refs. \cite{Ablowitz:96} and \cite{Bertola:16}. 
Such observations are also known to be related to the phenomenon of
higher modulation instability recently observed in optics
and in hydrodynamics
\cite{Hammani:11b,Erkintalo:11,Kimmoun:17,shaping_light_book_kibler},
wherein non-ideal excitation of first-order breathers (ABs, PS...)
or propagation losses were identified as perturbations that
introduce deviations from the expected nonlinear dynamics on
longer propagation distances (simply due the unstable nature
of NLS breathers). In particular, it was shown that higher
order modulation instability arises from perturbations on
breathers close to the PS limit, thus resulting in a nonlinear
superposition (i.e., complex arrangement) of several elementary breathers.

\begin{figure}[h]
\includegraphics[width=9.cm]{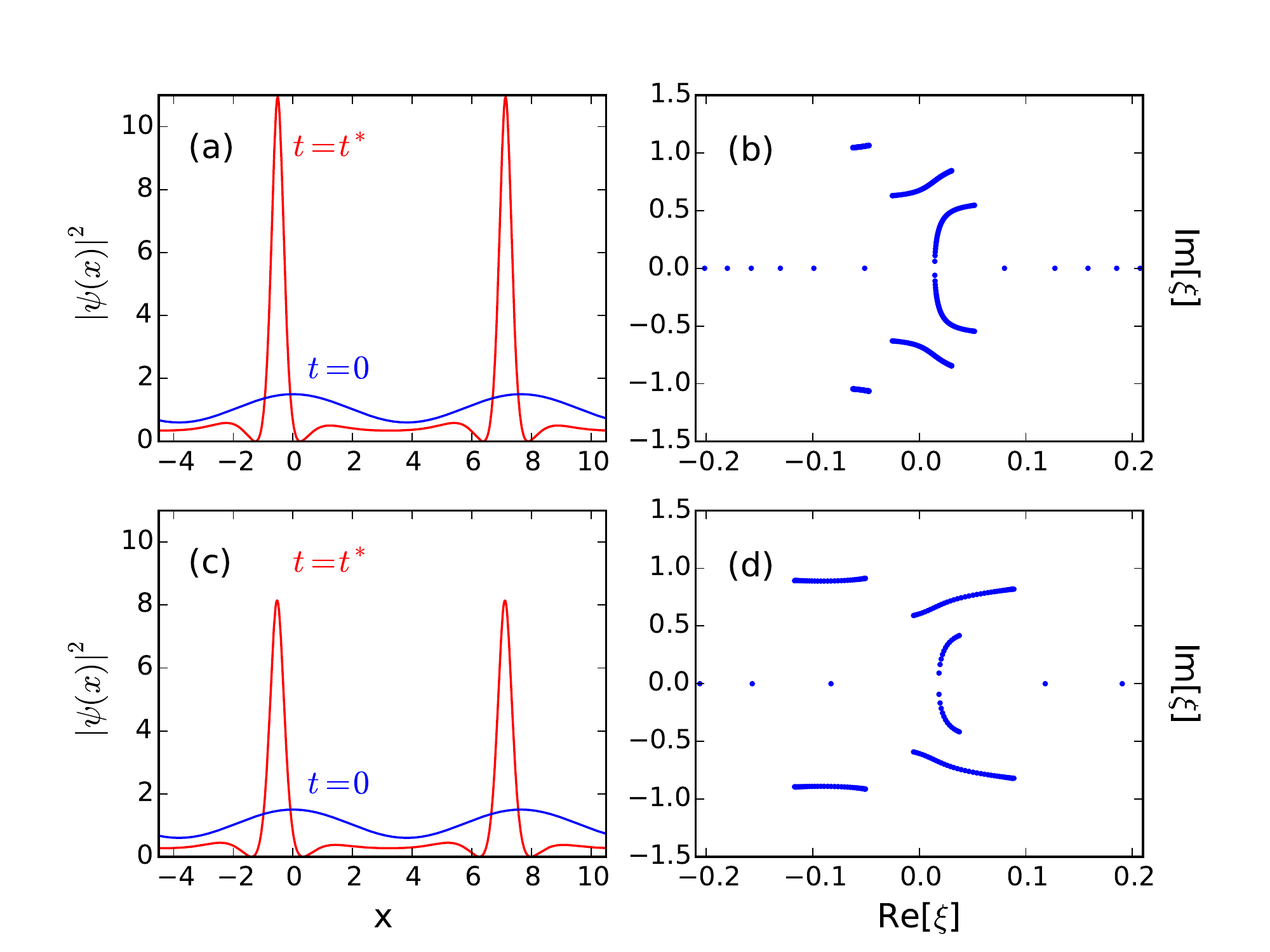}
\caption{(a) Numerical simulations of Eq. (\ref{nlse}) 
  with the initial condition given by Eq. (\ref{ci_opt})
  ($\epsilon=0.225$, $k_0=0.822$) and (b) corresponding IST spectrum. 
  (c) Numerical simulations of Eq. (\ref{nlse_damping}) 
  with the initial condition given by Eq. (\ref{ci_opt}).
  (d) IST spectrum of the coherent structure $\psi(x,t^*)$
  plotted with red lines in (c). The numerical values of the parameters
  correspond the optical
  fiber experiment reported in ref. \cite{Kibler:10} and their numerical values
  are $\epsilon=0.225$, $k_0=0.822$, $\kappa=0.076$.
  $t^{*}$ represent the times at which the breather structures reach
  their maximum intensities: (a) $t^{*}=1.192$, (c) $t^{*}=1.273$.
  The size $L=7.64$ of the box used for numerical IST analysis
  is equal to the period of the modulated wave ($L=2\pi/k_0$).
}
\label{fig:ist_nlse_opt_exp_2}
\end{figure}

Fig. 4(a) shows the result of the numerical
simulation of Eq. (\ref{nlse}) for a value of $\epsilon=0.225$ identical
the one typifying the optical fiber experiment reported
in ref. \cite{Kibler:10}. Fig. 4(b) represents the 
corresponding IST spectrum that displays five well-separated bands.
This shows that the generated structure remains a genus $4$
solution of the 1D-NLSE when the value of $\epsilon$ is increased from
$0.05$ to $0.225$. The peak intensity
of the generated breather structure reaches a value $\sim 11$ at
the time $t^*=1.192$. This value is greater than the peak
intensity of $9$ that can be reached by the PS
but it should be kept in mind that fiber losses have not
yet been taken into account in the numerical simulation. 

For a better quantitative description of the experiment 
reported in ref. \cite{Kibler:10}, fiber 
losses should be taken into account. To describe their influence, 
a linear damping term must be added to Eq. (\ref{nlse}) that becomes
\begin{equation}\label{nlse_damping}
  i \psi_t +  \psi_{xx} +  2 \,  |\psi|^2 \, \psi + i \, \kappa \, \psi=0.
\end{equation}
$\kappa \in \mathbb{R}^+$ is a damping parameter that
depends on the fiber attenuation coefficient, see Appendix \ref{appendixa}
for the correspondance between physical parameters and
dimensionless parameters. 
Fig. 4(c) shows the result of the numerical simulation
of Eq. (\ref{nlse_damping}) with the parameters $\epsilon=0.225$,  $k_0=0.822$,
$\kappa=0.076$ that characterize the 
optical fiber experiment reported in ref. \cite{Kibler:10}.
Now that the effect of dissipation is taken into account,
numerical simulations of the experiment show that the maximum
peak intensity reached by the breather structure is
$\sim 8.2$ at time $t^*=1.273$.
Moreover the core part of the intensity profile of the structure
numerically obtained can be well fitted by Eq. (\ref{PS}), like in
ref. \cite{Kibler:10}.

Now that dissipation is included in the 1D-NLSE, the wave
system can no longer be considered as being described by an integrable
equation. Rigorously speaking, the IST tools described in
Sec. \ref{ist_theory} can no longer be used to analyze
the generated structures. 
However, it should be emphasized that dissipation
is weak enough in the experiment ($\kappa=0.073$) to be considered
as a small perturbative effect. In these circumstances,
the fact that the optical power slightly decays all along the
optical fiber induces some slow modulation of the coherent structure.
If the IST analysis of the generated structure is made locally in time
(i.e. after some propagation distance inside the fiber) in the
presence of dissipation, we observe a gradual deformation of the
spectral bands. At a time $t\sim0$, the numerically
computed spectrum is very close to the one shown in Fig. 4(b)
but at time $t^*=1.273$, the spectral deformation is no longer negligible,
see Fig. 4(d). Dissipative effects break the integrability of the
wave equation, and the isospectrally condition (i. e. time-independence
of the IST spectrum) can no longer be satisfied in these circumstances.
However, it must be emphasized that dissipative effects remain of
perturbative nature and that their presence does not corrupt the
qualitative structure of the IST spectrum
in the sense that it is still composed of five bands. 

In conclusion, the numerical IST analysis of the experiment reported
in ref. \cite{Kibler:10} reveals that the initial condition used
in the experiment exhibits $5$ spectral bands, i.e. can be
more accurately approximated by a genus $4$ solution of the 1D-NLSE
at $t=0$, i.e. at the input end of the fiber.
Weak dissipative effects occurring inside the optical fiber
play a perturbative role and produce a slow variation
of the intensity profile and, therefore, of the spectral characteristics
of the generated solution. From the perspective of nonlinear
spectral analysis, the observed breather structure, therefore, represents a
slowly modulated genus $4$ solution of the 1D-NLSE with the modulation
of spectral characteristics  occurring due to weak dissipative effects. 

\subsection{Nonlinear spectral analysis of the experimental data}\label{sec:PS_opt_data_analysis}

In the optical fiber experiment reported in ref. \cite{Kibler:10},
the intensity and the phase of the optical signal have been measured over
several periods by using the so-called Frequency Resolved Optical
Gating (FROG) technique. The experimentally recorded signals
rescaled to dimensionless variables are plotted in red lines
in Fig. 5(a) and 5(b), see Appendix \ref{appendixa} for  
the correspondance between physical variables and
dimensionless variables. The blue lines in Fig. 5(a) and 5(b) represent
the intensity profile $|\psi(x)|^2=|\psi(x,t^*)|^2$
and the phase profile $\phi(x)=\phi(x,t^*)$ of the breather
structure ($\psi(x)=|\psi(x)|e^{i\phi(x)}$) computed from numerical
simulations of Eq. (\ref{nlse_damping}) with the dimensionless parameters
that characterize the optical fiber experiment, i. e. the
function $|\psi(x)|^2$ plotted with blue lines in Fig. 5(b)
is identical to the function $|\psi(x,t^*)|^2$ plotted in Fig. 4(c). 

Similar to ref. \cite{Kibler:10}, Fig. 5(b) shows that there
is a very good quantitative agreement
between the intensity profile that has been experimentally
recorded and the intensity profile that is computed from numerical
integration of Eq. (\ref{nlse_damping}).
However, the careful analysis of experimental data recorded in
ref. \cite{Kibler:10} reveals that the SFB structure that
is generated in the experiment present phase and intensity
profiles that are slowly modulated not only by fiber losses but also
by pertubative third-order dispersive effects occurring inside
the optical fiber. Indeed, 
the phase profile of SFB experimentally recorded
exhibits a marked asymmetry in the the region surrounding
the soliton core part of the observed SFB. A
phase jump of $\sim - \pi$ measured on the left-hand side of
the coherent structure is followed by another phase jump is
of $\sim -\pi$ on the
right-hand side of the coherent structure.  Such a pronounced
asymmetry is not found in the numerical simulation of
Eq. (\ref{nlse_damping}) that reveals a nearly symmetric phase profile with
a phase jump of $\sim +\pi$ followed by another phase jump of
$\sim - \pi$, see blue line in Fig. 5(a).
The asymmetry of the phase profile experimentally observed arises
from the occurrence of perturbative third-order dispersion terms
that are not included in Eq. (\ref{nlse_damping}). 
Inclusion of the third-order dispersion term in numerical simulations
reproduces quantitatively well the asymmetry of the 
phase profile experimentally observed without significantly
modifying the intensity profile
computed from numerical integration of Eq. (\ref{nlse_damping}).
This asymmetry in the phase jump simply arises from the fact that
one side of the breather has passed the maximal focusing point
before the other side (i.e., the phase tips over
$+\pi \leftrightarrow -\pi$) .

\begin{figure}[h]
\includegraphics[width=9.cm]{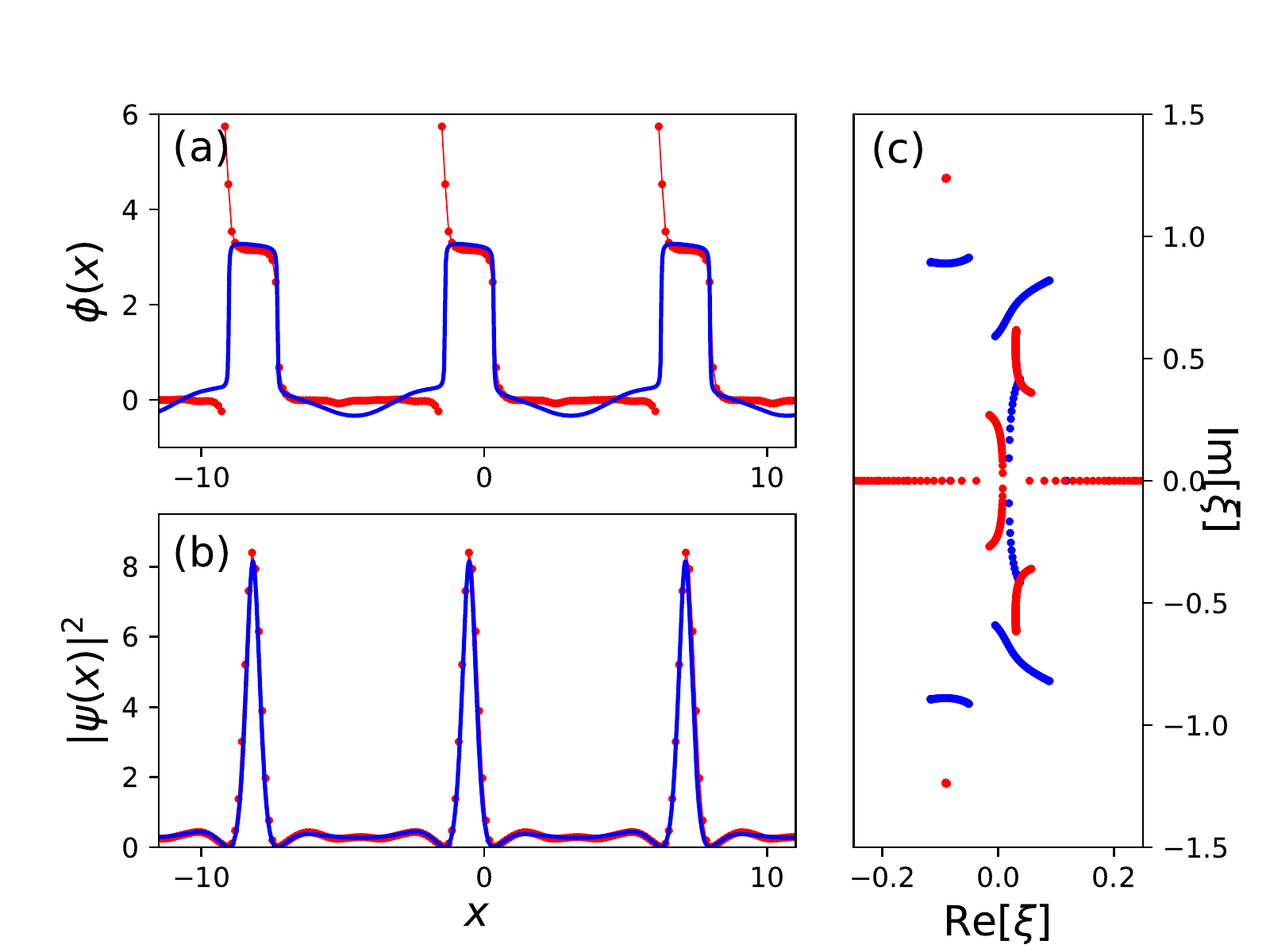}
\caption{(a) Phase profile (red line) and (b) intensity profile
  (red line) recorded in the optical fiber experiment reported
  in ref. \cite{Kibler:10}. The phase and intensity profiles
  plotted with blue lines in (a) and (b) are computed from the
  numerical simulation of Eq. (\ref{nlse_damping}) with parameters
  of the experiment ($\epsilon=0.225$, $k_0=0.822$, $\kappa=0.076$).
  (c) Spectral portraits computed from the optical signals recorded in the
  experiment (red dots) and from numerical simulations of 
  Eq. (\ref{nlse_damping}) (blue dots).
  }
\label{fig:ist_ps_opt}
\end{figure}

Fig. 5(c) shows the result of the nonlinear spectral analysis of
the experiment reported in ref. \cite{Kibler:10}. The spectrum
plotted in red dots is the spectrum of the optical signal
that has been experimentally recorded in ref. \cite{Kibler:10}.
The spectrum plotted in blue dots is the spectrum of the
coherent structure $\psi(x,t^*)$ computed from numerical
simulation of Eq. (\ref{nlse_damping}) with the parameters
of the experiments, i. e. the spectrum in blue dots in Fig. 5(c)
is identical to the spectrum in Fig. 4(d). Due to third-order dispersion,
the phase profiles are different in the experiment and in the numerical
simulation of Eq. (\ref{nlse_damping}).  
Therefore the two spectral
portraits in Fig. 4(c) are not identical. However, the nonlinear spectral
analysis of the experimental data reveals that the SFB structure
observed in the optical fiber experiment has a spectrum composed
of five bands which means that the observed SFB structure
has to be classified as a genus $4$ solution of the 1D-NLSE. 

In conclusion, the nonlinear spectral analysis of experimental
data reported in ref. \cite{Kibler:10} shows that the observed
structure can be more accurately described by a genus $4$ solution of the 1D-NLSE
having a spectral portrait composed of $5$ bands. 
Even though this solution has a mathematical nature that is different 
from the PS's one (a degenerate genus 2 solution of the 1D-NLSE), 
the SFB locally observed in optical fiber experiments retains some degree 
of proximity to the PS in the sense that it has similar properties 
of localization in space and time. 
The detailed nonlinear spectral analysis also reveals that 
weak dissipative effects and third-order dispersive effects
occurring inside the optical fiber play some perturbative roles 
by producing a slow modulation (with propagation distance)
of the spectral characteristics of the observed SFB. 

\section{Nonlinear spectral analysis of the
  Peregrine soliton observed in water tank experiments} \label{sec:PS_hydro}

In the hydrodynamic experiment reported in
ref. \cite{Chabchoub:11}, a breather structure is generated
from the motion of a flap inside a $15$ m $\times$ $1.6$ m
$\times$ $1.5$ m water wave tank with $1$ m water depth. 
The flap displacement was chosen to be proportional to the
amplitude of the PS to be generated, i.e
the flap motion is set to be directly proportional
to the water surface elevation that is determined from Eq. (\ref{PS}).
The generation of the initial condition is
made by chosing a time $t_0<0$ in Eq. (\ref{PS})
in such a way that the generated breather 
reaches its maximum amplitude near the end of the water tank.

\begin{figure}[h]
\includegraphics[width=9.cm]{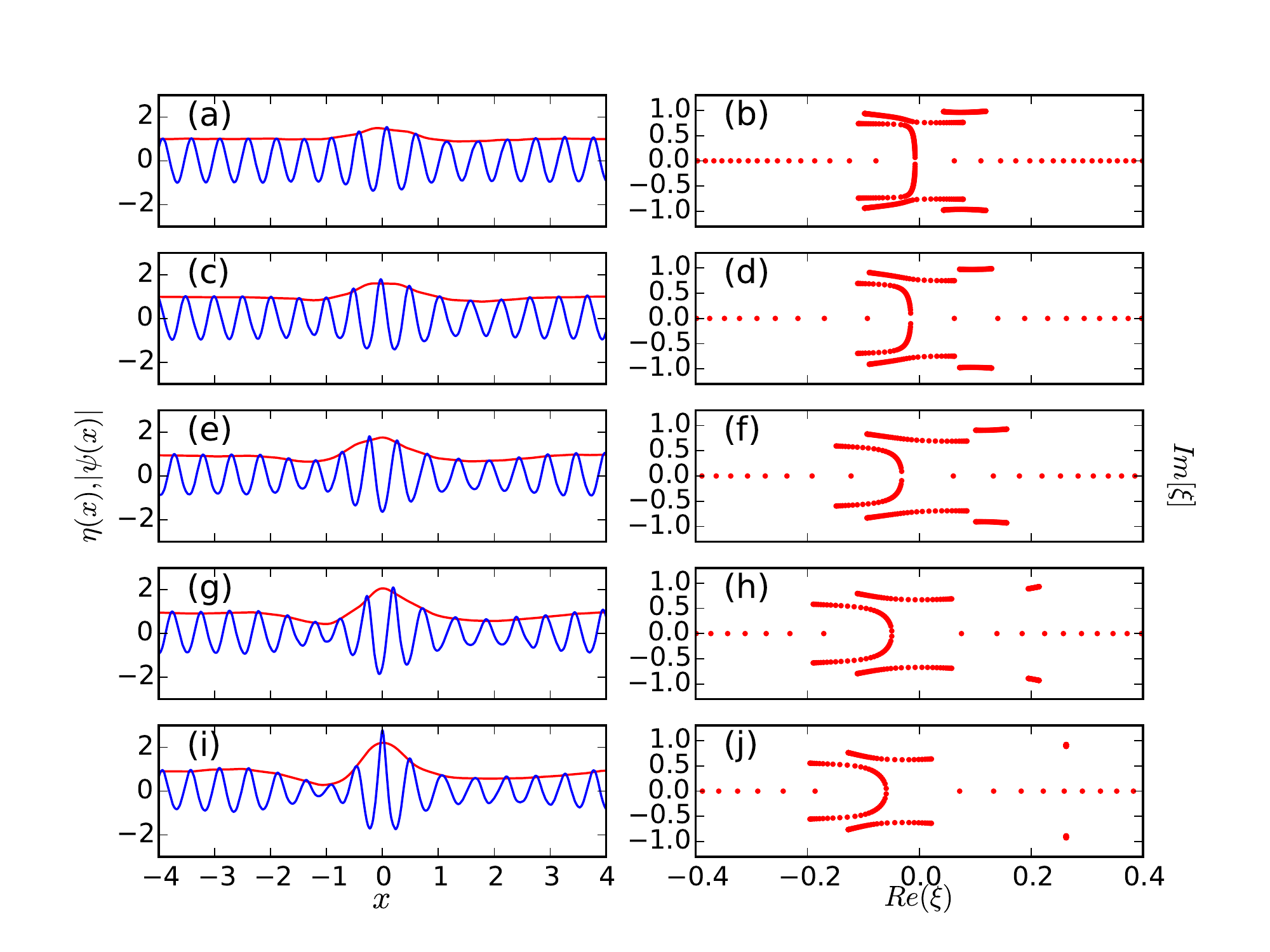}
\caption{Nonlinear spectral analysis of the experimental
  data recorded in the hydrodynamic experiment reported in
  ref. \cite{Chabchoub:11}. Left column: water elevation $\eta(x,t^*)$
  (blue lines) and modulus $|\psi(x,t^*)|$ (red lines) of the complex envelope
  of the wave packets. The evolution times $t^*$ and corresponding
  propagation distances
  $D$ inside the water tank are (a) $t^*=7.8 \times 10^{-3}$, $D=0.1$ m,
  (c) $t^*=0.1573$, $D=2.1$ m, (e) $t^*=0.3146$, $D=4.1$ m,
  (g) $t^*=0.4719$, $D=6.1$ m, (i) $t^*=0.629$, $D=8.1$ m.
  Right column: spectral portraits of the experimental signals.
  Computation of the spectral portraits is made with a numerical box
  having a size $L=8$.
}
\label{fig:ist_ps_hydro}
\end{figure}

In the experiment reported in
ref. \cite{Chabchoub:11}, the water waves that propagate inside the
tank are progressive Stokes waves having their complex envelopes $\psi$
that are slowly modulated in space and time. The water elevation $\eta$ 
is the real variable that is measured at several localized points
inside the water tank, see Fig. 3 of ref. \cite{Chabchoub:11} or
left column of Fig. 6. As discussed
in detail, e.g. in ref. \cite{Osborne2010nonlinear},
the complex envelope $\psi$ can be constructed from the measurement
of the free surface elevation $\eta$ by using the Hilbert transform.

In our analysis of the hydrodynamic data recorded
in ref. \cite{Chabchoub:11}, the signals recorded by several
capacitance gauges
are first rescaled to dimensionless form, see Appendix \ref{appendixb} for  
the correspondance between physical variables and
dimensionless variables. Then, the Hilbert transform is used to
determine the complex envelope $\psi(x,t^*)$ of the wave
at different times $t^*$ that correspond to the different
positions of the gauges inside the water tank. The normalized
surface elevation $\eta(x,t^*)$ and the corresponding envelope
$|\psi(x,t^*)|$ obtained by using this procedure are plotted
in the left column of Fig. 6.

\begin{figure}[h]
\includegraphics[width=9.cm]{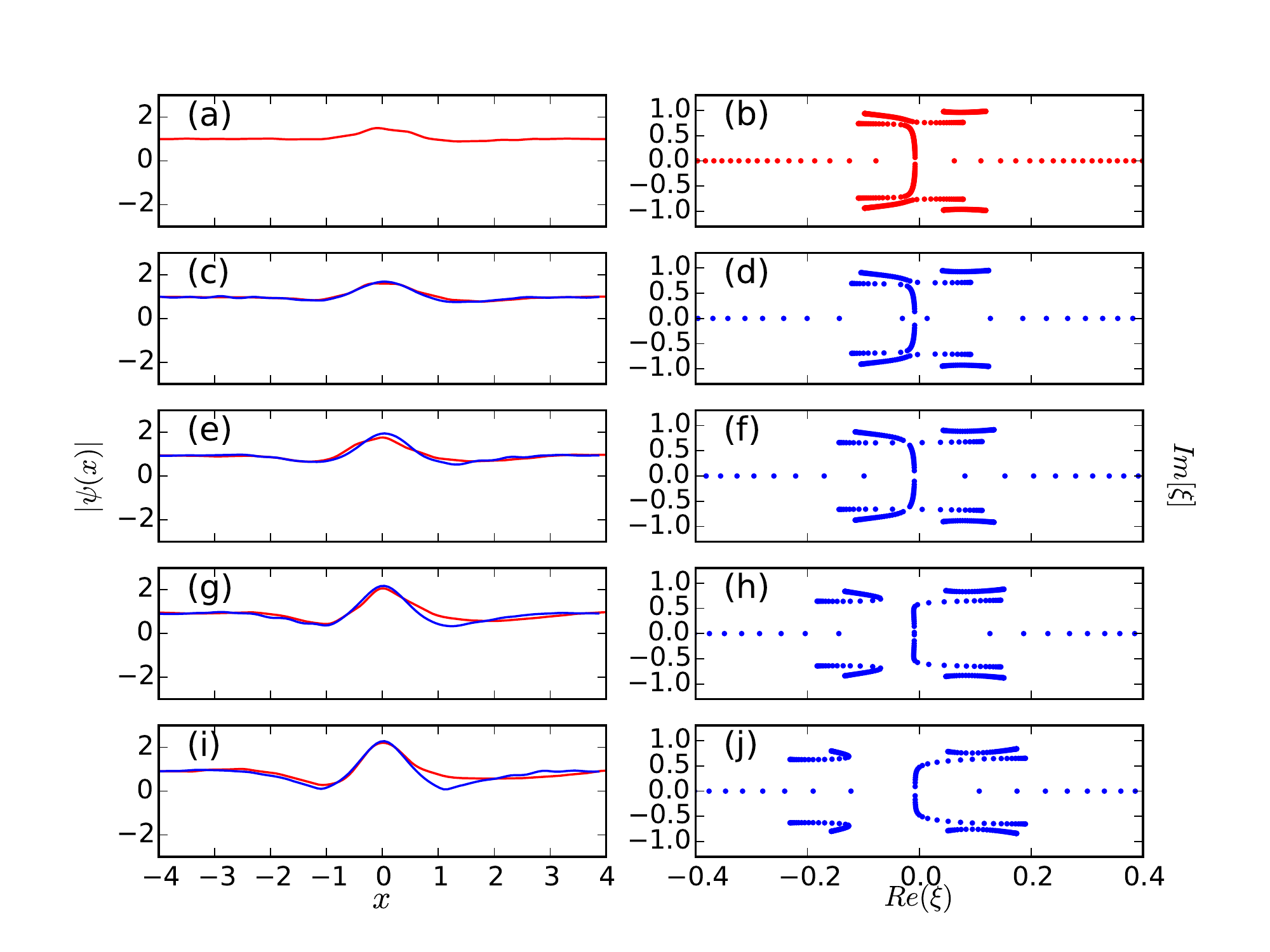}
\caption{Left column: Numerical simulations of Eq. (\ref{nlse_damping})
  while taking the signal measured by the first gauge (red line in (a))
  as initial condition. The red lines represent experimental signals
  and the blue lines represent the results of the numerical simulation
  of Eq. (\ref{nlse_damping}) with $\kappa=1.6 \times 10^{-2}$.
  The evolution times $t^*$ and the corresponding propagation
  distances $D$ inside the water tank are the same
  as in Fig. 6, i.e. 
  (a) $t^*=7.8 \times 10^{-3}$, $D=0.1$ m,
  (c) $t^*=0.1573$, $D=2.1$ m, (e) $t^*=0.3146$, $D=4.1$ m,
  (g) $t^*=0.4719$, $D=6.1$ m, (i) $t^*=0.629$, $D=8.1$ m.
  Right column: spectral portraits of the signals computed from 
  numerical simulation of Eq. (\ref{nlse_damping}).
  Computation of the spectral portraits is made with a numerical box
  having a size $L=8$.
    }
\label{fig:ist_ps_hydro_nlse}
\end{figure}

Contrary to the optical fiber experiment discussed in detail
in Sec. \ref{sec:PS_opt}, the hydrodynamic experiment
reported in ref. \cite{Chabchoub:11} presents the advantage
that the buildup of the breather structure can be observed
and measured. Nonlinear spectral analysis of the experimental
signals plotted in Fig. 6 reveals that the observed
structure represents a genus $4$ solution of the
1D-NLSE having its spectral properties
that are slowly modulated in time, see right column
in Fig. 6. The spectrum at the initial time (i.e. near the
wavemaker) is composed of five bands and it is relatively
confined around the vertical imaginary axis. It is
remarkable to observe that the nonlinear spectral analysis
evidences a slow gradual deformation of the five spectral bands
with time (i. e. with propagation distance along
the water tank). Fig. 6 indeed shows that the spectrum
is not time-independent but that it slowly spreads
along the real axis. Similar spectral behavior has been
evidenced in Sec. \ref{sec:PS_opt_numerical_analysis}
where dissipation has been shown to induce qualitatively
the same kind of spectral effect, cf. Fig. 4(b) and 4(d). 

These results suggest to conduct numerical simulations of Eq. (\ref{nlse_damping})
to investigate the influence of dissipation on the shape of the
spectral portraits. We have made numerical simulations
of Eq. (\ref{nlse_damping}) by taking an initial condition
that is given by the signal recorded by the
first capacitance gauge located only $10$ cm away from the wavemaker. 
The numerical value of the damping parameter $\kappa=1.6 \times 10^{-2}$
has been determined from
the measurement of the attenuation of the signals recorded
by all the other capacitance gauges, see Appendix \ref{appendixb}. 
Taking the signal recorded by the first gauge as initial
condition for the integration of Eq. (\ref{nlse_damping}),
the modulus of $\psi$ that is computed from the numerical simulation
is in good quantitative agreement with the envelope of the experimental
signals, see left column in Fig. 7. Only a slight asymmetry
measured in the experiment is not found at long evolution time
(i.e. far from the wavemaker) in the numerical simulation of
Eq. (\ref{nlse_damping}), see Fig. 7(g)(i).

Performing the nonlinear spectral analysis of the function
$\psi(x,t^*)$ computed from numerical simulations at evolution times 
$t^*$ corresponding to discrete positions at which the signals have been
recorded in the experiment, we observe a spreading of the
spectral portrait around the vertical imaginary axis that has already
been evidenced in Fig. 6 in the nonlinear spectral analysis
of the experimental data. This slow modulation of the
spectral characteristics of the analyzed structure
shows that dissipation plays a measurable role in the 
experiment reported in ref. \cite{Chabchoub:11}. 
Note that recent works have shown that dissipation
plays a significant influence in the buildup dynamics of
breather solutions of the 1D-NLSE \cite{Kimmoun:16,Kimmoun:17}.

Note that there is a reasonably good quantitative agreement
between spectral portaits plotted in Fig. 6(d)(f) and in
Fig. 7(d)(f). This shows that the dynamics of the wavepacket
is correctly described by Eq. (\ref{nlse_damping}) until
$t^*=0.314$ (i.e. until the wave packet reaches the gauge that
is located $5.10$ m away from the wavemaker). Beyond this evolution
time, the spectra computed from the analysis of the experimental
data and from the numerical simulation of Eq. (\ref{nlse_damping})
become of qualitatively different natures, compare Fig. 6(h)(j) and
Fig. 7(h)(j). This means that wave propagation is not
ruled by Eq. (\ref{nlse_damping}) for evolution times
greater than $0.314$ (i.e. for propagation distances greater
than $5.10$ m). Nonlinear spectral analysis thus reveals
that higher order effects play some non-negligible role 
in the buildup dynamics of the breather structure
for the chosen carrier parameters that has a significant
steepness of $0.1163$. With this value of the steepness, higher-order effects are
to be expected given the significant focusing of the wave
field \cite{Shemer:02,Islas:11}. Note that their contribution can be
reduced, when choosing smaller carrier steepness values in accordance
with weakly nonlinear theory \cite{Chabchoub:12c}.

\section{Discussion and Conclusion}\label{conclusion}

In this paper, we have analyzed the data recorded in 
an optical fiber experiment \cite{Kibler:10} and in a hydrodynamic 
experiment \cite{Chabchoub:11}
that have reported the observation of SFB having properties 
of localization in space and time similar to those of the PS. 
Our approach is based  on the integrable nature of the 1D-NLSE
that governs at leading order the propagation of the optical 
and hydrodynamical waves in these two experiments. 
Solving the ZS spatial problem with periodic boundary conditions
by using Floquet spectral theory, we have computed 
the spectral (IST) portraits of the PSs observed in the two 
experiments, thus obtaining their nonlinear spectral signature.  

Contrary to linear (Fourier) spectral analysis that 
mostly provides spectra having some universal triangular shape 
whatever the exact nature of the SFB under consideration
\cite{Kibler:10,Frisquet:13,Frisquet:14,Kibler:15,Akhmediev:11a}, we obtain 
spectral signatures having some shapes that strongly depend
on the exact phase and intensity profiles of the examined SFB. 
These spectral portraits can be interpreted within the framework 
of the so-called finite gap theory (or periodic IST). 
In particular, the number $g+1$ of bands composing the 
nonlinear spectrum determines the genus $g$ of the solution
that parametrizes the complexity of the space-time 
evolution of the considered solution. Note that phase portraits
recently used e. g. in ref. \cite{Kimmoun:16,Mussot:18}
provide some other useful geometric representation of the
dynamics of breather solutions of the 1D-NLSE.

The nonlinear spectral analysis of the SFB observed 
both in the optical fiber experiment and in the water tank
experiment reveals that they represent general genus $4$ solutions 
of the 1D-NLSE that may exhibit a space-time evolution
that is much more complex than a simpler degenerate genus $2$
solution like PS or ABs,
as shown in the recent experiments reported in ref.
\cite{Hammani:11b,Erkintalo:11,shaping_light_book_kibler}.
Moreover, nonlinear spectral analysis reveals 
that the spectrum of the observed SFB slowly changes 
with the propagation distance, thus confirming a clear
evidence of the influence of perturbative higher order 
effects like dissipation in the experiments \cite{Kimmoun:16,Kimmoun:17}.  

As discussed in detail in ref. \cite{Islas:05,Calini:12}, the size of the 
gaps between the endpoints of the spectral branches 
determines the proximity to some homoclinic solutions of the 
1D-NLSE. The smaller is the size of these gaps, the 
closer is the observed solution to some homoclinic solution 
associated with some high peak intensity. 
An appropriate (and non trivial) choice of the initial phases of 
the finite-gap solution would permit to reduce the size of 
the gaps, thus approaching homoclinic solutions. 

From a more practical perspective, nonlinear spectral 
analysis could possibly be used as a predictive tool for the 
prediction of the space-time evolution of coherent (soliton- or breather-like)
structures emerging in random wave systems ruled by 
integrable equations \cite{Walczak:15,Akhmediev:11b,Cousins:16}.
The measurement of the size of the 
spectral bands provides some information about the maximum
amplitude that can be possibly reached by the coherent structure 
under consideration. As discussed in ref. \cite{Islas:05,Calini:12},
the measurement of 
the size of the spectral gaps indicates the proximity to homoclinic 
orbits and is therefore related to the probability of emergence 
of a rogue wave.

\appendix

\section{Relation between physical variables and
  dimensionless variables in the optical fiber
  experiment}\label{appendixa} 

In the optical fiber experiment reported in ref. \cite{Kibler:10},
light propagation is governed by the dimensional 1D-NLSE
\begin{equation}\label{nlse_phys}
  i \frac{\partial A }{\partial z} - \frac{\beta_2}{2} \frac{\partial^2 A}{\partial T^2} + \gamma |A|^2 A + i \frac{\alpha}{2} A =0 
\end{equation}
where third-order dispersion effects are not taken into account.
$A(z,T)$ is the slowly-varying complex envelope of the electric
field propagating inside the core of the single-mode fiber. $z$ represents
the longitudinal coordinate measuring the propagation distance
along the fiber and $T$ is the time measured in 
the retarded frame moving with light pulses. $\beta_2$
and $\gamma$ represent the second-order dispersion coefficient
and the Kerr coefficient of the optical fiber, respectively.
For the fiber that was used in ref. \cite{Kibler:10}, these
parameters are $\beta_2=-8.85 \, \times \, 10^{-28}$ s$^2$ m$^{-1}$
and $\gamma=0.01$ W$^{-1}$ m$^{-1}$. $\alpha$ is the fiber power attenuation
coefficient that is equal to $2.3 \, \times \, 10^{-4}$  m$^{-1}$
($\alpha=1$ dB/km). 

Considering that the optical input power is $P_0$, the normalized
optical field is defined by $\psi=A/\sqrt{P_0}$.
Taking a definition commonly adopted in nonlinear fiber
optics, the nonlinear length is given by $L_{NL}=1/(\gamma P_0)$.
In the optical fiber experiment of ref. \cite{Kibler:10},
the input power $P_0$ is $300$ mW and the nonlinear length is
of $333$ m.
Introducing the dimensionless time and space variables $t=2\gamma P_0 \, z$
and $x=\sqrt{\gamma P_0 / |\beta_2|} \, T$ together with the damping
parameter $\kappa=\alpha/(\gamma P_0)$, Eq. (\ref{nlse_phys})
takes the following dimensionless form
\begin{equation}\label{nlse_damping2}
  i \psi_t +  \psi_{xx} +  2 \,  |\psi|^2 \, \psi + i \, \kappa \, \psi=0.
\end{equation}
In the optical fiber experiment of ref. \cite{Kibler:10},
the damping parameter is
$\kappa=7.6 \times 10^{-2}$. 

The spatial frequency $k_0$ of the dimensionless modulated field 
that is used as initial condition (see Eq. (\ref{ci_opt})) is given by 
\begin{equation}\label{modulation_frequency}
  k_0=2 \pi \Delta f \sqrt{\frac{|\beta_2|}{\gamma P_0}}
\end{equation}
where $\Delta f=241$ GHz is the frequency detuning between
the two lasers used in the experiment. With the numerical values
of the physical parameters used in the experiment, the numerical
value of $k_0$ is $0.822$.

\section{Relation between physical variables and
  dimensionless variables in the hydrodynamic
  experiment}\label{appendixb} 

In the hydrodynamic experiment reported in ref. \cite{Kibler:10},
the evolution of the one-dimensional deep-water packets can
be described by the following equation \cite{Kimmoun:16}: 
\begin{equation}\label{nlse_phys_hydro}
  i \left(\frac{\partial a }{\partial z} + \frac{1}{c_g} \frac{\partial a }{\partial \tau} \right) - \frac{1}{g} \frac{\partial^2 a}{\partial \tau^2} -k_0^3 |a|^2 a = i \Gamma a . 
\end{equation}
$\tau$ and $z$ represent the time and the longitudinal coordinate
measuring the propagation distance alond the 1D-flume, respectively.
$g$ denotes the gravitational acceleration and $k_0$ is the wave
number. $k_0$ is linked to the pulsation $\omega_0$ of the
carrier wave by the dispersion relation of the linear deep
water wave theory $\omega_0=\sqrt{g\,k_0}$. $c_g=\frac{\omega_0}{2 k_0}$
represents the group velocity of the wave packets. The surface
elevation $\eta(z,\tau)$ of water is given by $\eta(z,\tau)=\Re(a(z,\tau)
\exp[i(k_0 z-\omega_0 t)])$. $\Gamma$ represents the dissipation rate.
Introducing the amplitude of the carrier wave $a_0$, the steepness
is defined by $\epsilon=a_0 k_0$.

Introducing the dimensionless time and space variables
$t=-\frac{1}{2}a_0^2k_0^3z$ and
$x=\frac{\sqrt{2}a_0k_0\omega_0}{2}\left(\tau-\frac{z}{c_g}\right)$
together with the normalized complex envelope
of the elevation field $\psi=a/a_0$, Eq. (\ref{nlse_phys_hydro})
becomes
\begin{equation}\label{nlse_damping3}
  i \psi_t +  \psi_{xx} +  2 \,  |\psi|^2 \, \psi + i \, \kappa \, \psi=0
\end{equation}
where $\kappa=\frac{2\Gamma}{a_0^2 k_0^3}$. In the hydrodynamic
experiment reported in ref. \cite{Chabchoub:11},
the numerical values of $a_0$, $k_0$, $\omega_0$ are
$0.01$ m, $11.63$ m$^{-1}$, $10.7$ s$^{-1}$, respectively.
The dissipation rate $\Gamma$ measured from experimental
signals is $\Gamma=1.26 \times 10^{-2}$ m$^{-1}$ and the
value of the damping parameter is $\kappa=1.6 \times 10^{-2}$.

\begin{acknowledgments}
This work has been partially supported by the Agence Nationale de la
Recherche through the LABEX CEMPI project (ANR-11-LABX-0007)   and by the Ministry of
Higher Education and Research, Hauts-de-France Regional Council and
European Regional Development Fund (ERDF) through the Contrat de
Projets Etat-R\'egion (CPER Photonics for Society P4S)  and by the Centre National de la Recherche Scientifique (CNRS) through the project MICRO TURBU and
the funding program ``Emergence 2017''.  The work of GE was partially supported by EPSRC grant
EP/R00515X/1.  The authors thank G. Roberti for helpful comments and J. P. Flament,
F. R\'eal and V. Vallet from Laboratoire PhLAM for technical assistance
with computer ressources.
\end{acknowledgments}


\begin{thebibliography}{92}%
\makeatletter
\providecommand \@ifxundefined [1]{%
 \@ifx{#1\undefined}
}%
\providecommand \@ifnum [1]{%
 \ifnum #1\expandafter \@firstoftwo
 \else \expandafter \@secondoftwo
 \fi
}%
\providecommand \@ifx [1]{%
 \ifx #1\expandafter \@firstoftwo
 \else \expandafter \@secondoftwo
 \fi
}%
\providecommand \natexlab [1]{#1}%
\providecommand \enquote  [1]{``#1''}%
\providecommand \bibnamefont  [1]{#1}%
\providecommand \bibfnamefont [1]{#1}%
\providecommand \citenamefont [1]{#1}%
\providecommand \href@noop [0]{\@secondoftwo}%
\providecommand \href [0]{\begingroup \@sanitize@url \@href}%
\providecommand \@href[1]{\@@startlink{#1}\@@href}%
\providecommand \@@href[1]{\endgroup#1\@@endlink}%
\providecommand \@sanitize@url [0]{\catcode `\\12\catcode `\$12\catcode
  `\&12\catcode `\#12\catcode `\^12\catcode `\_12\catcode `\%12\relax}%
\providecommand \@@startlink[1]{}%
\providecommand \@@endlink[0]{}%
\providecommand \url  [0]{\begingroup\@sanitize@url \@url }%
\providecommand \@url [1]{\endgroup\@href {#1}{\urlprefix }}%
\providecommand \urlprefix  [0]{URL }%
\providecommand \Eprint [0]{\href }%
\providecommand \doibase [0]{http://dx.doi.org/}%
\providecommand \selectlanguage [0]{\@gobble}%
\providecommand \bibinfo  [0]{\@secondoftwo}%
\providecommand \bibfield  [0]{\@secondoftwo}%
\providecommand \translation [1]{[#1]}%
\providecommand \BibitemOpen [0]{}%
\providecommand \bibitemStop [0]{}%
\providecommand \bibitemNoStop [0]{.\EOS\space}%
\providecommand \EOS [0]{\spacefactor3000\relax}%
\providecommand \BibitemShut  [1]{\csname bibitem#1\endcsname}%
\let\auto@bib@innerbib\@empty
\bibitem [{\citenamefont {Kibler}\ \emph {et~al.}(2010)\citenamefont {Kibler},
  \citenamefont {Fatome}, \citenamefont {Finot}, \citenamefont {Millot},
  \citenamefont {Dias}, \citenamefont {Genty}, \citenamefont {Akhmediev},\ and\
  \citenamefont {Dudley}}]{Kibler:10}%
  \BibitemOpen
  \bibfield  {author} {\bibinfo {author} {\bibfnamefont {B.}~\bibnamefont
  {Kibler}}, \bibinfo {author} {\bibfnamefont {J.}~\bibnamefont {Fatome}},
  \bibinfo {author} {\bibfnamefont {C.}~\bibnamefont {Finot}}, \bibinfo
  {author} {\bibfnamefont {G.}~\bibnamefont {Millot}}, \bibinfo {author}
  {\bibfnamefont {F.}~\bibnamefont {Dias}}, \bibinfo {author} {\bibfnamefont
  {G.}~\bibnamefont {Genty}}, \bibinfo {author} {\bibfnamefont
  {N.}~\bibnamefont {Akhmediev}}, \ and\ \bibinfo {author} {\bibfnamefont
  {J.~M.}\ \bibnamefont {Dudley}},\ }\href@noop {} {\bibfield  {journal}
  {\bibinfo  {journal} {Nature Physics}\ }\textbf {\bibinfo {volume} {6}},\
  \bibinfo {pages} {790} (\bibinfo {year} {2010})}\BibitemShut {NoStop}%
\bibitem [{\citenamefont {Chabchoub}\ \emph {et~al.}(2011)\citenamefont
  {Chabchoub}, \citenamefont {Hoffmann},\ and\ \citenamefont
  {Akhmediev}}]{Chabchoub:11}%
  \BibitemOpen
  \bibfield  {author} {\bibinfo {author} {\bibfnamefont {A.}~\bibnamefont
  {Chabchoub}}, \bibinfo {author} {\bibfnamefont {N.~P.}\ \bibnamefont
  {Hoffmann}}, \ and\ \bibinfo {author} {\bibfnamefont {N.}~\bibnamefont
  {Akhmediev}},\ }\href@noop {} {\bibfield  {journal} {\bibinfo  {journal}
  {Phys. Rev. Lett.}\ }\textbf {\bibinfo {volume} {106}},\ \bibinfo {pages}
  {204502} (\bibinfo {year} {2011})}\BibitemShut {NoStop}%
\bibitem [{\citenamefont {Yang}(2010)}]{yang2010nonlinear}%
  \BibitemOpen
  \bibfield  {author} {\bibinfo {author} {\bibfnamefont {J.}~\bibnamefont
  {Yang}},\ }\href {https://books.google.fr/books?id=ACbbjcRQQvUC} {\emph
  {\bibinfo {title} {Nonlinear Waves in Integrable and Non-integrable
  Systems}}},\ Mathematical Modeling and Computation\ (\bibinfo  {publisher}
  {Society for Industrial and Applied Mathematics},\ \bibinfo {year}
  {2010})\BibitemShut {NoStop}%
\bibitem [{\citenamefont {Akhmediev}\ and\ \citenamefont
  {Ankiewicz}(1997)}]{akhmediev1997solitons}%
  \BibitemOpen
  \bibfield  {author} {\bibinfo {author} {\bibfnamefont {N.~N.}\ \bibnamefont
  {Akhmediev}}\ and\ \bibinfo {author} {\bibfnamefont {A.}~\bibnamefont
  {Ankiewicz}},\ }\href@noop {} {\emph {\bibinfo {title} {Solitons: nonlinear
  pulses and beams}}}\ (\bibinfo  {publisher} {Chapman \& Hall},\ \bibinfo
  {year} {1997})\BibitemShut {NoStop}%
\bibitem [{\citenamefont {Ablowitz}(2011)}]{ablowitz2011nonlinear}%
  \BibitemOpen
  \bibfield  {author} {\bibinfo {author} {\bibfnamefont {M.~J.}\ \bibnamefont
  {Ablowitz}},\ }\href@noop {} {\emph {\bibinfo {title} {Nonlinear dispersive
  waves: asymptotic analysis and solitons}}},\ Vol.~\bibinfo {volume} {47}\
  (\bibinfo  {publisher} {Cambridge University Press},\ \bibinfo {year}
  {2011})\BibitemShut {NoStop}%
\bibitem [{\citenamefont {Zakharov}\ and\ \citenamefont
  {Shabat}(1972)}]{Zakharov:72}%
  \BibitemOpen
  \bibfield  {author} {\bibinfo {author} {\bibfnamefont {V.~E.}\ \bibnamefont
  {Zakharov}}\ and\ \bibinfo {author} {\bibfnamefont {A.~B.}\ \bibnamefont
  {Shabat}},\ }\href@noop {} {\bibfield  {journal} {\bibinfo  {journal} {Sov.
  Phys.--JETP}\ }\textbf {\bibinfo {volume} {34}},\ \bibinfo {pages} {62}
  (\bibinfo {year} {1972})}\BibitemShut {NoStop}%
\bibitem [{\citenamefont {Ablowitz}\ \emph {et~al.}(1974)\citenamefont
  {Ablowitz}, \citenamefont {Kaup}, \citenamefont {Newell},\ and\ \citenamefont
  {Segur}}]{Ablowitz:74}%
  \BibitemOpen
  \bibfield  {author} {\bibinfo {author} {\bibfnamefont {M.~J.}\ \bibnamefont
  {Ablowitz}}, \bibinfo {author} {\bibfnamefont {D.~J.}\ \bibnamefont {Kaup}},
  \bibinfo {author} {\bibfnamefont {A.~C.}\ \bibnamefont {Newell}}, \ and\
  \bibinfo {author} {\bibfnamefont {H.}~\bibnamefont {Segur}},\ }\href
  {\doibase 10.1002/sapm1974534249} {\bibfield  {journal} {\bibinfo  {journal}
  {Studies in Applied Mathematics}\ }\textbf {\bibinfo {volume} {53}},\
  \bibinfo {pages} {249} (\bibinfo {year} {1974})}\BibitemShut {NoStop}%
\bibitem [{\citenamefont {Zabusky}\ and\ \citenamefont
  {Kruskal}(1965)}]{Zabusky:65}%
  \BibitemOpen
  \bibfield  {author} {\bibinfo {author} {\bibfnamefont {N.~J.}\ \bibnamefont
  {Zabusky}}\ and\ \bibinfo {author} {\bibfnamefont {M.~D.}\ \bibnamefont
  {Kruskal}},\ }\href {\doibase 10.1103/PhysRevLett.15.240} {\bibfield
  {journal} {\bibinfo  {journal} {Phys. Rev. Lett.}\ }\textbf {\bibinfo
  {volume} {15}},\ \bibinfo {pages} {240} (\bibinfo {year} {1965})}\BibitemShut
  {NoStop}%
\bibitem [{\citenamefont {Trillo}\ \emph {et~al.}(2016)\citenamefont {Trillo},
  \citenamefont {Deng}, \citenamefont {Biondini}, \citenamefont {Klein},
  \citenamefont {Clauss}, \citenamefont {Chabchoub},\ and\ \citenamefont
  {Onorato}}]{Trillo:16}%
  \BibitemOpen
  \bibfield  {author} {\bibinfo {author} {\bibfnamefont {S.}~\bibnamefont
  {Trillo}}, \bibinfo {author} {\bibfnamefont {G.}~\bibnamefont {Deng}},
  \bibinfo {author} {\bibfnamefont {G.}~\bibnamefont {Biondini}}, \bibinfo
  {author} {\bibfnamefont {M.}~\bibnamefont {Klein}}, \bibinfo {author}
  {\bibfnamefont {G.~F.}\ \bibnamefont {Clauss}}, \bibinfo {author}
  {\bibfnamefont {A.}~\bibnamefont {Chabchoub}}, \ and\ \bibinfo {author}
  {\bibfnamefont {M.}~\bibnamefont {Onorato}},\ }\href {\doibase
  10.1103/PhysRevLett.117.144102} {\bibfield  {journal} {\bibinfo  {journal}
  {Phys. Rev. Lett.}\ }\textbf {\bibinfo {volume} {117}},\ \bibinfo {pages}
  {144102} (\bibinfo {year} {2016})}\BibitemShut {NoStop}%
\bibitem [{\citenamefont {Kuznetsov}(1977)}]{kuznetsov1977solitons}%
  \BibitemOpen
  \bibfield  {author} {\bibinfo {author} {\bibfnamefont {E.}~\bibnamefont
  {Kuznetsov}},\ }in\ \href@noop {} {\emph {\bibinfo {booktitle} {Akademiia
  Nauk SSSR Doklady}}},\ Vol.\ \bibinfo {volume} {236}\ (\bibinfo {year}
  {1977})\ pp.\ \bibinfo {pages} {575--577}\BibitemShut {NoStop}%
\bibitem [{\citenamefont {Ma}(1979)}]{ma1979perturbed}%
  \BibitemOpen
  \bibfield  {author} {\bibinfo {author} {\bibfnamefont {Y.}~\bibnamefont
  {Ma}},\ }\href@noop {} {\bibfield  {journal} {\bibinfo  {journal} {Studies in
  Applied Mathematics}\ }\textbf {\bibinfo {volume} {60}},\ \bibinfo {pages}
  {43} (\bibinfo {year} {1979})}\BibitemShut {NoStop}%
\bibitem [{\citenamefont {Peregrine}(1983)}]{peregrine1983water}%
  \BibitemOpen
  \bibfield  {author} {\bibinfo {author} {\bibfnamefont {D.}~\bibnamefont
  {Peregrine}},\ }\href@noop {} {\bibfield  {journal} {\bibinfo  {journal} {J.
  Austral. Math. Soc. Ser. B}\ }\textbf {\bibinfo {volume} {25}},\ \bibinfo
  {pages} {16} (\bibinfo {year} {1983})}\BibitemShut {NoStop}%
\bibitem [{\citenamefont {Akhmediev}\ \emph {et~al.}(1985)\citenamefont
  {Akhmediev}, \citenamefont {Eleonskii},\ and\ \citenamefont
  {Kulagin}}]{akhmediev1985generation}%
  \BibitemOpen
  \bibfield  {author} {\bibinfo {author} {\bibfnamefont {N.}~\bibnamefont
  {Akhmediev}}, \bibinfo {author} {\bibfnamefont {V.}~\bibnamefont
  {Eleonskii}}, \ and\ \bibinfo {author} {\bibfnamefont {N.}~\bibnamefont
  {Kulagin}},\ }\href@noop {} {\bibfield  {journal} {\bibinfo  {journal} {Sov.
  Phys. JETP}\ }\textbf {\bibinfo {volume} {62}},\ \bibinfo {pages} {894}
  (\bibinfo {year} {1985})}\BibitemShut {NoStop}%
\bibitem [{\citenamefont {Kibler}\ \emph {et~al.}(2012)\citenamefont {Kibler},
  \citenamefont {Fatome}, \citenamefont {Finot}, \citenamefont {Millot},
  \citenamefont {Genty}, \citenamefont {Wetzel}, \citenamefont {Akhmediev},
  \citenamefont {Dias},\ and\ \citenamefont {Dudley}}]{Kibler:12}%
  \BibitemOpen
  \bibfield  {author} {\bibinfo {author} {\bibfnamefont {B.}~\bibnamefont
  {Kibler}}, \bibinfo {author} {\bibfnamefont {J.}~\bibnamefont {Fatome}},
  \bibinfo {author} {\bibfnamefont {C.}~\bibnamefont {Finot}}, \bibinfo
  {author} {\bibfnamefont {G.}~\bibnamefont {Millot}}, \bibinfo {author}
  {\bibfnamefont {G.}~\bibnamefont {Genty}}, \bibinfo {author} {\bibfnamefont
  {B.}~\bibnamefont {Wetzel}}, \bibinfo {author} {\bibfnamefont
  {N.}~\bibnamefont {Akhmediev}}, \bibinfo {author} {\bibfnamefont
  {F.}~\bibnamefont {Dias}}, \ and\ \bibinfo {author} {\bibfnamefont {J.~M.}\
  \bibnamefont {Dudley}},\ }\href@noop {} {\bibfield  {journal} {\bibinfo
  {journal} {Scientific Reports}\ }\textbf {\bibinfo {volume} {2}} (\bibinfo
  {year} {2012})}\BibitemShut {NoStop}%
\bibitem [{\citenamefont {Hammani}\ \emph
  {et~al.}(2011{\natexlab{a}})\citenamefont {Hammani}, \citenamefont {Wetzel},
  \citenamefont {Kibler}, \citenamefont {Fatome}, \citenamefont {Finot},
  \citenamefont {Millot}, \citenamefont {Akhmediev},\ and\ \citenamefont
  {Dudley}}]{Hammani:11}%
  \BibitemOpen
  \bibfield  {author} {\bibinfo {author} {\bibfnamefont {K.}~\bibnamefont
  {Hammani}}, \bibinfo {author} {\bibfnamefont {B.}~\bibnamefont {Wetzel}},
  \bibinfo {author} {\bibfnamefont {B.}~\bibnamefont {Kibler}}, \bibinfo
  {author} {\bibfnamefont {J.}~\bibnamefont {Fatome}}, \bibinfo {author}
  {\bibfnamefont {C.}~\bibnamefont {Finot}}, \bibinfo {author} {\bibfnamefont
  {G.}~\bibnamefont {Millot}}, \bibinfo {author} {\bibfnamefont
  {N.}~\bibnamefont {Akhmediev}}, \ and\ \bibinfo {author} {\bibfnamefont
  {J.~M.}\ \bibnamefont {Dudley}},\ }\href@noop {} {\bibfield  {journal}
  {\bibinfo  {journal} {Optics letters}\ }\textbf {\bibinfo {volume} {36}},\
  \bibinfo {pages} {2140} (\bibinfo {year} {2011}{\natexlab{a}})}\BibitemShut
  {NoStop}%
\bibitem [{\citenamefont {Chabchoub}\ \emph {et~al.}(2014)\citenamefont
  {Chabchoub}, \citenamefont {Kibler}, \citenamefont {Dudley},\ and\
  \citenamefont {Akhmediev}}]{chabchoub:14}%
  \BibitemOpen
  \bibfield  {author} {\bibinfo {author} {\bibfnamefont {A.}~\bibnamefont
  {Chabchoub}}, \bibinfo {author} {\bibfnamefont {B.}~\bibnamefont {Kibler}},
  \bibinfo {author} {\bibfnamefont {J.~M.}\ \bibnamefont {Dudley}}, \ and\
  \bibinfo {author} {\bibfnamefont {N.}~\bibnamefont {Akhmediev}},\ }\href@noop
  {} {\bibfield  {journal} {\bibinfo  {journal} {Philosophical Transactions of
  the Royal Society of London A: Mathematical, Physical and Engineering
  Sciences}\ }\textbf {\bibinfo {volume} {372}} (\bibinfo {year}
  {2014})}\BibitemShut {NoStop}%
\bibitem [{\citenamefont {Onorato}\ \emph {et~al.}(2013)\citenamefont
  {Onorato}, \citenamefont {Residori}, \citenamefont {Bortolozzo},
  \citenamefont {Montina},\ and\ \citenamefont {Arecchi}}]{Onorato:13}%
  \BibitemOpen
  \bibfield  {author} {\bibinfo {author} {\bibfnamefont {M.}~\bibnamefont
  {Onorato}}, \bibinfo {author} {\bibfnamefont {S.}~\bibnamefont {Residori}},
  \bibinfo {author} {\bibfnamefont {U.}~\bibnamefont {Bortolozzo}}, \bibinfo
  {author} {\bibfnamefont {A.}~\bibnamefont {Montina}}, \ and\ \bibinfo
  {author} {\bibfnamefont {F.}~\bibnamefont {Arecchi}},\ }\href@noop {}
  {\bibfield  {journal} {\bibinfo  {journal} {Physics Reports}\ }\textbf
  {\bibinfo {volume} {528}},\ \bibinfo {pages} {47 } (\bibinfo {year}
  {2013})}\BibitemShut {NoStop}%
\bibitem [{\citenamefont {Shrira}\ and\ \citenamefont
  {Geogjaev}(2010)}]{Shrira:10}%
  \BibitemOpen
  \bibfield  {author} {\bibinfo {author} {\bibfnamefont {V.~I.}\ \bibnamefont
  {Shrira}}\ and\ \bibinfo {author} {\bibfnamefont {V.~V.}\ \bibnamefont
  {Geogjaev}},\ }\href {\doibase 10.1007/s10665-009-9347-2} {\bibfield
  {journal} {\bibinfo  {journal} {Journal of Engineering Mathematics}\ }\textbf
  {\bibinfo {volume} {67}},\ \bibinfo {pages} {11} (\bibinfo {year}
  {2010})}\BibitemShut {NoStop}%
\bibitem [{\citenamefont {Dubrovin}\ \emph {et~al.}(2009)\citenamefont
  {Dubrovin}, \citenamefont {Grava},\ and\ \citenamefont
  {Klein}}]{Dubrovin:09}%
  \BibitemOpen
  \bibfield  {author} {\bibinfo {author} {\bibfnamefont {B.}~\bibnamefont
  {Dubrovin}}, \bibinfo {author} {\bibfnamefont {T.}~\bibnamefont {Grava}}, \
  and\ \bibinfo {author} {\bibfnamefont {C.}~\bibnamefont {Klein}},\ }\href
  {\doibase 10.1007/s00332-008-9025-y} {\bibfield  {journal} {\bibinfo
  {journal} {Journal of Nonlinear Science}\ }\textbf {\bibinfo {volume} {19}},\
  \bibinfo {pages} {57} (\bibinfo {year} {2009})}\BibitemShut {NoStop}%
\bibitem [{\citenamefont {Bertola}\ and\ \citenamefont
  {Tovbis}(2013)}]{Bertola:13}%
  \BibitemOpen
  \bibfield  {author} {\bibinfo {author} {\bibfnamefont {M.}~\bibnamefont
  {Bertola}}\ and\ \bibinfo {author} {\bibfnamefont {A.}~\bibnamefont
  {Tovbis}},\ }\href {\doibase 10.1002/cpa.21445} {\bibfield  {journal}
  {\bibinfo  {journal} {Communications on Pure and Applied Mathematics}\
  }\textbf {\bibinfo {volume} {66}},\ \bibinfo {pages} {678} (\bibinfo {year}
  {2013})}\BibitemShut {NoStop}%
\bibitem [{\citenamefont {Tikan}\ \emph {et~al.}(2017)\citenamefont {Tikan},
  \citenamefont {Billet}, \citenamefont {El}, \citenamefont {Tovbis},
  \citenamefont {Bertola}, \citenamefont {Sylvestre}, \citenamefont {Gustave},
  \citenamefont {Randoux}, \citenamefont {Genty}, \citenamefont {Suret},\ and\
  \citenamefont {Dudley}}]{Tikan:17}%
  \BibitemOpen
  \bibfield  {author} {\bibinfo {author} {\bibfnamefont {A.}~\bibnamefont
  {Tikan}}, \bibinfo {author} {\bibfnamefont {C.}~\bibnamefont {Billet}},
  \bibinfo {author} {\bibfnamefont {G.}~\bibnamefont {El}}, \bibinfo {author}
  {\bibfnamefont {A.}~\bibnamefont {Tovbis}}, \bibinfo {author} {\bibfnamefont
  {M.}~\bibnamefont {Bertola}}, \bibinfo {author} {\bibfnamefont
  {T.}~\bibnamefont {Sylvestre}}, \bibinfo {author} {\bibfnamefont
  {F.}~\bibnamefont {Gustave}}, \bibinfo {author} {\bibfnamefont
  {S.}~\bibnamefont {Randoux}}, \bibinfo {author} {\bibfnamefont
  {G.}~\bibnamefont {Genty}}, \bibinfo {author} {\bibfnamefont
  {P.}~\bibnamefont {Suret}}, \ and\ \bibinfo {author} {\bibfnamefont {J.~M.}\
  \bibnamefont {Dudley}},\ }\href {\doibase 10.1103/PhysRevLett.119.033901}
  {\bibfield  {journal} {\bibinfo  {journal} {Phys. Rev. Lett.}\ }\textbf
  {\bibinfo {volume} {119}},\ \bibinfo {pages} {033901} (\bibinfo {year}
  {2017})}\BibitemShut {NoStop}%
\bibitem [{\citenamefont {Akhmediev}\ \emph
  {et~al.}(2009{\natexlab{a}})\citenamefont {Akhmediev}, \citenamefont
  {Ankiewicz},\ and\ \citenamefont {Taki}}]{Akhmediev:09}%
  \BibitemOpen
  \bibfield  {author} {\bibinfo {author} {\bibfnamefont {N.}~\bibnamefont
  {Akhmediev}}, \bibinfo {author} {\bibfnamefont {A.}~\bibnamefont
  {Ankiewicz}}, \ and\ \bibinfo {author} {\bibfnamefont {M.}~\bibnamefont
  {Taki}},\ }\href@noop {} {\bibfield  {journal} {\bibinfo  {journal} {Physics
  Letters A}\ }\textbf {\bibinfo {volume} {373}},\ \bibinfo {pages} {675 }
  (\bibinfo {year} {2009}{\natexlab{a}})}\BibitemShut {NoStop}%
\bibitem [{\citenamefont {Akhmediev}\ \emph
  {et~al.}(2009{\natexlab{b}})\citenamefont {Akhmediev}, \citenamefont
  {Ankiewicz},\ and\ \citenamefont {Soto-Crespo}}]{Akhmediev:09c}%
  \BibitemOpen
  \bibfield  {author} {\bibinfo {author} {\bibfnamefont {N.}~\bibnamefont
  {Akhmediev}}, \bibinfo {author} {\bibfnamefont {A.}~\bibnamefont
  {Ankiewicz}}, \ and\ \bibinfo {author} {\bibfnamefont {J.~M.}\ \bibnamefont
  {Soto-Crespo}},\ }\href {\doibase 10.1103/PhysRevE.80.026601} {\bibfield
  {journal} {\bibinfo  {journal} {Phys. Rev. E}\ }\textbf {\bibinfo {volume}
  {80}},\ \bibinfo {pages} {026601} (\bibinfo {year}
  {2009}{\natexlab{b}})}\BibitemShut {NoStop}%
\bibitem [{\citenamefont {Frisquet}\ \emph {et~al.}(2013)\citenamefont
  {Frisquet}, \citenamefont {Kibler},\ and\ \citenamefont
  {Millot}}]{Frisquet:13}%
  \BibitemOpen
  \bibfield  {author} {\bibinfo {author} {\bibfnamefont {B.}~\bibnamefont
  {Frisquet}}, \bibinfo {author} {\bibfnamefont {B.}~\bibnamefont {Kibler}}, \
  and\ \bibinfo {author} {\bibfnamefont {G.}~\bibnamefont {Millot}},\
  }\href@noop {} {\bibfield  {journal} {\bibinfo  {journal} {Physical Review
  X}\ }\textbf {\bibinfo {volume} {3}},\ \bibinfo {pages} {041032} (\bibinfo
  {year} {2013})}\BibitemShut {NoStop}%
\bibitem [{\citenamefont {Chabchoub}\ \emph
  {et~al.}(2012{\natexlab{a}})\citenamefont {Chabchoub}, \citenamefont
  {Hoffmann}, \citenamefont {Onorato},\ and\ \citenamefont
  {Akhmediev}}]{Chabchoub:12}%
  \BibitemOpen
  \bibfield  {author} {\bibinfo {author} {\bibfnamefont {A.}~\bibnamefont
  {Chabchoub}}, \bibinfo {author} {\bibfnamefont {N.}~\bibnamefont {Hoffmann}},
  \bibinfo {author} {\bibfnamefont {M.}~\bibnamefont {Onorato}}, \ and\
  \bibinfo {author} {\bibfnamefont {N.}~\bibnamefont {Akhmediev}},\ }\href@noop
  {} {\bibfield  {journal} {\bibinfo  {journal} {Physical Review X}\ }\textbf
  {\bibinfo {volume} {2}},\ \bibinfo {pages} {011015} (\bibinfo {year}
  {2012}{\natexlab{a}})}\BibitemShut {NoStop}%
\bibitem [{\citenamefont {Chabchoub}\ \emph
  {et~al.}(2012{\natexlab{b}})\citenamefont {Chabchoub}, \citenamefont
  {Hoffmann}, \citenamefont {Onorato}, \citenamefont {Slunyaev}, \citenamefont
  {Sergeeva}, \citenamefont {Pelinovsky},\ and\ \citenamefont
  {Akhmediev}}]{Chabchoub:12b}%
  \BibitemOpen
  \bibfield  {author} {\bibinfo {author} {\bibfnamefont {A.}~\bibnamefont
  {Chabchoub}}, \bibinfo {author} {\bibfnamefont {N.}~\bibnamefont {Hoffmann}},
  \bibinfo {author} {\bibfnamefont {M.}~\bibnamefont {Onorato}}, \bibinfo
  {author} {\bibfnamefont {A.}~\bibnamefont {Slunyaev}}, \bibinfo {author}
  {\bibfnamefont {A.}~\bibnamefont {Sergeeva}}, \bibinfo {author}
  {\bibfnamefont {E.}~\bibnamefont {Pelinovsky}}, \ and\ \bibinfo {author}
  {\bibfnamefont {N.}~\bibnamefont {Akhmediev}},\ }\href {\doibase
  10.1103/PhysRevE.86.056601} {\bibfield  {journal} {\bibinfo  {journal} {Phys.
  Rev. E}\ }\textbf {\bibinfo {volume} {86}},\ \bibinfo {pages} {056601}
  (\bibinfo {year} {2012}{\natexlab{b}})}\BibitemShut {NoStop}%
\bibitem [{\citenamefont {Zakharov}\ and\ \citenamefont
  {Gelash}(2013)}]{Zakharov:13}%
  \BibitemOpen
  \bibfield  {author} {\bibinfo {author} {\bibfnamefont {V.~E.}\ \bibnamefont
  {Zakharov}}\ and\ \bibinfo {author} {\bibfnamefont {A.~A.}\ \bibnamefont
  {Gelash}},\ }\href@noop {} {\bibfield  {journal} {\bibinfo  {journal} {Phys.
  Rev. Lett.}\ }\textbf {\bibinfo {volume} {111}},\ \bibinfo {pages} {054101}
  (\bibinfo {year} {2013})}\BibitemShut {NoStop}%
\bibitem [{\citenamefont {Kibler}\ \emph {et~al.}(2015)\citenamefont {Kibler},
  \citenamefont {Chabchoub}, \citenamefont {Gelash}, \citenamefont
  {Akhmediev},\ and\ \citenamefont {Zakharov}}]{Kibler:15}%
  \BibitemOpen
  \bibfield  {author} {\bibinfo {author} {\bibfnamefont {B.}~\bibnamefont
  {Kibler}}, \bibinfo {author} {\bibfnamefont {A.}~\bibnamefont {Chabchoub}},
  \bibinfo {author} {\bibfnamefont {A.}~\bibnamefont {Gelash}}, \bibinfo
  {author} {\bibfnamefont {N.}~\bibnamefont {Akhmediev}}, \ and\ \bibinfo
  {author} {\bibfnamefont {V.~E.}\ \bibnamefont {Zakharov}},\ }\href {\doibase
  10.1103/PhysRevX.5.041026} {\bibfield  {journal} {\bibinfo  {journal} {Phys.
  Rev. X}\ }\textbf {\bibinfo {volume} {5}},\ \bibinfo {pages} {041026}
  (\bibinfo {year} {2015})}\BibitemShut {NoStop}%
\bibitem [{\citenamefont {Biondini}\ and\ \citenamefont
  {Mantzavinos}(2016)}]{Biondini:16a}%
  \BibitemOpen
  \bibfield  {author} {\bibinfo {author} {\bibfnamefont {G.}~\bibnamefont
  {Biondini}}\ and\ \bibinfo {author} {\bibfnamefont {D.}~\bibnamefont
  {Mantzavinos}},\ }\href@noop {} {\bibfield  {journal} {\bibinfo  {journal}
  {Phys. Rev. Lett.}\ }\textbf {\bibinfo {volume} {116}},\ \bibinfo {pages}
  {043902} (\bibinfo {year} {2016})}\BibitemShut {NoStop}%
\bibitem [{\citenamefont {Kraych}\ \emph {et~al.}(2018)\citenamefont {Kraych},
  \citenamefont {Suret}, \citenamefont {El},\ and\ \citenamefont
  {Randoux}}]{Kraych:18}%
  \BibitemOpen
  \bibfield  {author} {\bibinfo {author} {\bibfnamefont {A.~E.}\ \bibnamefont
  {Kraych}}, \bibinfo {author} {\bibfnamefont {P.}~\bibnamefont {Suret}},
  \bibinfo {author} {\bibfnamefont {G.}~\bibnamefont {El}}, \ and\ \bibinfo
  {author} {\bibfnamefont {S.}~\bibnamefont {Randoux}},\ }\href@noop {}
  {\bibfield  {journal} {\bibinfo  {journal} {Arxiv preprint arXiv:1805.05074}\
  } (\bibinfo {year} {2018})}\BibitemShut {NoStop}%
\bibitem [{\citenamefont {Zakharov}(2009)}]{Zakharov:09}%
  \BibitemOpen
  \bibfield  {author} {\bibinfo {author} {\bibfnamefont {V.~E.}\ \bibnamefont
  {Zakharov}},\ }\href@noop {} {\bibfield  {journal} {\bibinfo  {journal}
  {Studies in Applied Mathematics}\ }\textbf {\bibinfo {volume} {122}},\
  \bibinfo {pages} {219} (\bibinfo {year} {2009})}\BibitemShut {NoStop}%
\bibitem [{\citenamefont {Randoux}\ \emph {et~al.}(2014)\citenamefont
  {Randoux}, \citenamefont {Walczak}, \citenamefont {Onorato},\ and\
  \citenamefont {Suret}}]{Randoux:14}%
  \BibitemOpen
  \bibfield  {author} {\bibinfo {author} {\bibfnamefont {S.}~\bibnamefont
  {Randoux}}, \bibinfo {author} {\bibfnamefont {P.}~\bibnamefont {Walczak}},
  \bibinfo {author} {\bibfnamefont {M.}~\bibnamefont {Onorato}}, \ and\
  \bibinfo {author} {\bibfnamefont {P.}~\bibnamefont {Suret}},\ }\href@noop {}
  {\bibfield  {journal} {\bibinfo  {journal} {Phys. Rev. Lett.}\ }\textbf
  {\bibinfo {volume} {113}},\ \bibinfo {pages} {113902} (\bibinfo {year}
  {2014})}\BibitemShut {NoStop}%
\bibitem [{\citenamefont {Walczak}\ \emph {et~al.}(2015)\citenamefont
  {Walczak}, \citenamefont {Randoux},\ and\ \citenamefont
  {Suret}}]{Walczak:15}%
  \BibitemOpen
  \bibfield  {author} {\bibinfo {author} {\bibfnamefont {P.}~\bibnamefont
  {Walczak}}, \bibinfo {author} {\bibfnamefont {S.}~\bibnamefont {Randoux}}, \
  and\ \bibinfo {author} {\bibfnamefont {P.}~\bibnamefont {Suret}},\
  }\href@noop {} {\bibfield  {journal} {\bibinfo  {journal} {Phys. Rev. Lett.}\
  }\textbf {\bibinfo {volume} {114}},\ \bibinfo {pages} {143903} (\bibinfo
  {year} {2015})}\BibitemShut {NoStop}%
\bibitem [{\citenamefont {Agafontsev}\ and\ \citenamefont
  {Zakharov}(2015)}]{Agafontsev:15}%
  \BibitemOpen
  \bibfield  {author} {\bibinfo {author} {\bibfnamefont {D.~S.}\ \bibnamefont
  {Agafontsev}}\ and\ \bibinfo {author} {\bibfnamefont {V.~E.}\ \bibnamefont
  {Zakharov}},\ }\href {http://stacks.iop.org/0951-7715/28/i=8/a=2791}
  {\bibfield  {journal} {\bibinfo  {journal} {Nonlinearity}\ }\textbf {\bibinfo
  {volume} {28}},\ \bibinfo {pages} {2791} (\bibinfo {year}
  {2015})}\BibitemShut {NoStop}%
\bibitem [{\citenamefont {Randoux}\ \emph
  {et~al.}(2016{\natexlab{a}})\citenamefont {Randoux}, \citenamefont {Walczak},
  \citenamefont {Onorato},\ and\ \citenamefont {Suret}}]{Randoux:16}%
  \BibitemOpen
  \bibfield  {author} {\bibinfo {author} {\bibfnamefont {S.}~\bibnamefont
  {Randoux}}, \bibinfo {author} {\bibfnamefont {P.}~\bibnamefont {Walczak}},
  \bibinfo {author} {\bibfnamefont {M.}~\bibnamefont {Onorato}}, \ and\
  \bibinfo {author} {\bibfnamefont {P.}~\bibnamefont {Suret}},\ }\href@noop {}
  {\bibfield  {journal} {\bibinfo  {journal} {Physica D: Nonlinear Phenomena}\
  }\textbf {\bibinfo {volume} {333}},\ \bibinfo {pages} {323} (\bibinfo {year}
  {2016}{\natexlab{a}})}\BibitemShut {NoStop}%
\bibitem [{\citenamefont {Randoux}\ \emph {et~al.}(2017)\citenamefont
  {Randoux}, \citenamefont {Gustave}, \citenamefont {Suret},\ and\
  \citenamefont {El}}]{Randoux:17}%
  \BibitemOpen
  \bibfield  {author} {\bibinfo {author} {\bibfnamefont {S.}~\bibnamefont
  {Randoux}}, \bibinfo {author} {\bibfnamefont {F.}~\bibnamefont {Gustave}},
  \bibinfo {author} {\bibfnamefont {P.}~\bibnamefont {Suret}}, \ and\ \bibinfo
  {author} {\bibfnamefont {G.}~\bibnamefont {El}},\ }\href {\doibase
  10.1103/PhysRevLett.118.233901} {\bibfield  {journal} {\bibinfo  {journal}
  {Phys. Rev. Lett.}\ }\textbf {\bibinfo {volume} {118}},\ \bibinfo {pages}
  {233901} (\bibinfo {year} {2017})}\BibitemShut {NoStop}%
\bibitem [{\citenamefont {Soto-Crespo}\ \emph {et~al.}(2016)\citenamefont
  {Soto-Crespo}, \citenamefont {Devine},\ and\ \citenamefont
  {Akhmediev}}]{Soto:16}%
  \BibitemOpen
  \bibfield  {author} {\bibinfo {author} {\bibfnamefont {J.~M.}\ \bibnamefont
  {Soto-Crespo}}, \bibinfo {author} {\bibfnamefont {N.}~\bibnamefont {Devine}},
  \ and\ \bibinfo {author} {\bibfnamefont {N.}~\bibnamefont {Akhmediev}},\
  }\href {\doibase 10.1103/PhysRevLett.116.103901} {\bibfield  {journal}
  {\bibinfo  {journal} {Phys. Rev. Lett.}\ }\textbf {\bibinfo {volume} {116}},\
  \bibinfo {pages} {103901} (\bibinfo {year} {2016})}\BibitemShut {NoStop}%
\bibitem [{\citenamefont {Akhmediev}\ \emph {et~al.}(2016)\citenamefont
  {Akhmediev}, \citenamefont {Soto-Crespo},\ and\ \citenamefont
  {Devine}}]{Akhmediev:16}%
  \BibitemOpen
  \bibfield  {author} {\bibinfo {author} {\bibfnamefont {N.}~\bibnamefont
  {Akhmediev}}, \bibinfo {author} {\bibfnamefont {J.~M.}\ \bibnamefont
  {Soto-Crespo}}, \ and\ \bibinfo {author} {\bibfnamefont {N.}~\bibnamefont
  {Devine}},\ }\href {\doibase 10.1103/PhysRevE.94.022212} {\bibfield
  {journal} {\bibinfo  {journal} {Phys. Rev. E}\ }\textbf {\bibinfo {volume}
  {94}},\ \bibinfo {pages} {022212} (\bibinfo {year} {2016})}\BibitemShut
  {NoStop}%
\bibitem [{\citenamefont {Dudley}\ \emph {et~al.}(2014)\citenamefont {Dudley},
  \citenamefont {Dias}, \citenamefont {Erkintalo},\ and\ \citenamefont
  {Genty}}]{Dudley:14}%
  \BibitemOpen
  \bibfield  {author} {\bibinfo {author} {\bibfnamefont {J.~M.}\ \bibnamefont
  {Dudley}}, \bibinfo {author} {\bibfnamefont {F.}~\bibnamefont {Dias}},
  \bibinfo {author} {\bibfnamefont {M.}~\bibnamefont {Erkintalo}}, \ and\
  \bibinfo {author} {\bibfnamefont {G.}~\bibnamefont {Genty}},\ }\href@noop {}
  {\bibfield  {journal} {\bibinfo  {journal} {Nat. Photon.}\ }\textbf {\bibinfo
  {volume} {8}},\ \bibinfo {pages} {755} (\bibinfo {year} {2014})}\BibitemShut
  {NoStop}%
\bibitem [{\citenamefont {Akhmediev}\ \emph
  {et~al.}(2009{\natexlab{c}})\citenamefont {Akhmediev}, \citenamefont
  {Soto-Crespo},\ and\ \citenamefont {Ankiewicz}}]{Akhmediev:09b}%
  \BibitemOpen
  \bibfield  {author} {\bibinfo {author} {\bibfnamefont {N.}~\bibnamefont
  {Akhmediev}}, \bibinfo {author} {\bibfnamefont {J.}~\bibnamefont
  {Soto-Crespo}}, \ and\ \bibinfo {author} {\bibfnamefont {A.}~\bibnamefont
  {Ankiewicz}},\ }\href@noop {} {\bibfield  {journal} {\bibinfo  {journal}
  {Physics Letters A}\ }\textbf {\bibinfo {volume} {373}},\ \bibinfo {pages}
  {2137 } (\bibinfo {year} {2009}{\natexlab{c}})}\BibitemShut {NoStop}%
\bibitem [{\citenamefont {Toenger}\ \emph {et~al.}(2015)\citenamefont
  {Toenger}, \citenamefont {Godin}, \citenamefont {Billet}, \citenamefont
  {Dias}, \citenamefont {Erkintalo}, \citenamefont {Genty},\ and\ \citenamefont
  {Dudley}}]{Toenger:15}%
  \BibitemOpen
  \bibfield  {author} {\bibinfo {author} {\bibfnamefont {S.}~\bibnamefont
  {Toenger}}, \bibinfo {author} {\bibfnamefont {T.}~\bibnamefont {Godin}},
  \bibinfo {author} {\bibfnamefont {C.}~\bibnamefont {Billet}}, \bibinfo
  {author} {\bibfnamefont {F.}~\bibnamefont {Dias}}, \bibinfo {author}
  {\bibfnamefont {M.}~\bibnamefont {Erkintalo}}, \bibinfo {author}
  {\bibfnamefont {G.}~\bibnamefont {Genty}}, \ and\ \bibinfo {author}
  {\bibfnamefont {J.~M.}\ \bibnamefont {Dudley}},\ }\href@noop {} {\bibfield
  {journal} {\bibinfo  {journal} {Sci. Rep.}\ }\textbf {\bibinfo {volume}
  {5}},\ \bibinfo {pages} {10380} (\bibinfo {year} {2015})}\BibitemShut
  {NoStop}%
\bibitem [{\citenamefont {Suret}\ \emph {et~al.}(2016)\citenamefont {Suret},
  \citenamefont {El~Koussaifi}, \citenamefont {Tikan}, \citenamefont {Evain},
  \citenamefont {Randoux}, \citenamefont {Szwaj},\ and\ \citenamefont
  {Bielawski}}]{Suret:16}%
  \BibitemOpen
  \bibfield  {author} {\bibinfo {author} {\bibfnamefont {P.}~\bibnamefont
  {Suret}}, \bibinfo {author} {\bibfnamefont {R.}~\bibnamefont {El~Koussaifi}},
  \bibinfo {author} {\bibfnamefont {A.}~\bibnamefont {Tikan}}, \bibinfo
  {author} {\bibfnamefont {C.}~\bibnamefont {Evain}}, \bibinfo {author}
  {\bibfnamefont {S.}~\bibnamefont {Randoux}}, \bibinfo {author} {\bibfnamefont
  {C.}~\bibnamefont {Szwaj}}, \ and\ \bibinfo {author} {\bibfnamefont
  {S.}~\bibnamefont {Bielawski}},\ }\href@noop {} {\bibfield  {journal}
  {\bibinfo  {journal} {Nature Communications}\ }\textbf {\bibinfo {volume}
  {7}},\ \bibinfo {pages} {13136} (\bibinfo {year} {2016})}\BibitemShut
  {NoStop}%
\bibitem [{\citenamefont {N\"arhi}\ \emph {et~al.}(2016)\citenamefont
  {N\"arhi}, \citenamefont {Wetzel}, \citenamefont {Billet}, \citenamefont
  {Toenger}, \citenamefont {Sylvestre}, \citenamefont {Merolla}, \citenamefont
  {Morandotti}, \citenamefont {Dias}, \citenamefont {Genty},\ and\
  \citenamefont {Dudley}}]{Nahri:16}%
  \BibitemOpen
  \bibfield  {author} {\bibinfo {author} {\bibfnamefont {M.}~\bibnamefont
  {N\"arhi}}, \bibinfo {author} {\bibfnamefont {B.}~\bibnamefont {Wetzel}},
  \bibinfo {author} {\bibfnamefont {C.}~\bibnamefont {Billet}}, \bibinfo
  {author} {\bibfnamefont {S.}~\bibnamefont {Toenger}}, \bibinfo {author}
  {\bibfnamefont {T.}~\bibnamefont {Sylvestre}}, \bibinfo {author}
  {\bibfnamefont {J.-M.}\ \bibnamefont {Merolla}}, \bibinfo {author}
  {\bibfnamefont {R.}~\bibnamefont {Morandotti}}, \bibinfo {author}
  {\bibfnamefont {F.}~\bibnamefont {Dias}}, \bibinfo {author} {\bibfnamefont
  {G.}~\bibnamefont {Genty}}, \ and\ \bibinfo {author} {\bibfnamefont {J.~M.}\
  \bibnamefont {Dudley}},\ }\href@noop {} {\bibfield  {journal} {\bibinfo
  {journal} {Nature Communications}\ }\textbf {\bibinfo {volume} {7}},\
  \bibinfo {pages} {13675} (\bibinfo {year} {2016})}\BibitemShut {NoStop}%
\bibitem [{\citenamefont {Tikan}\ \emph {et~al.}(2018)\citenamefont {Tikan},
  \citenamefont {Bielawski}, \citenamefont {Szwaj}, \citenamefont {Randoux},\
  and\ \citenamefont {Suret}}]{Tikan:18}%
  \BibitemOpen
  \bibfield  {author} {\bibinfo {author} {\bibfnamefont {A.}~\bibnamefont
  {Tikan}}, \bibinfo {author} {\bibfnamefont {S.}~\bibnamefont {Bielawski}},
  \bibinfo {author} {\bibfnamefont {C.}~\bibnamefont {Szwaj}}, \bibinfo
  {author} {\bibfnamefont {S.}~\bibnamefont {Randoux}}, \ and\ \bibinfo
  {author} {\bibfnamefont {P.}~\bibnamefont {Suret}},\ }\href {\doibase
  10.1038/s41566-018-0113-8} {\bibfield  {journal} {\bibinfo  {journal} {Nature
  Photonics}\ }\textbf {\bibinfo {volume} {12}},\ \bibinfo {pages} {228}
  (\bibinfo {year} {2018})}\BibitemShut {NoStop}%
\bibitem [{\citenamefont {Osborne}(2010)}]{Osborne2010nonlinear}%
  \BibitemOpen
  \bibfield  {author} {\bibinfo {author} {\bibfnamefont {A.}~\bibnamefont
  {Osborne}},\ }\href@noop {} {\emph {\bibinfo {title} {{Nonlinear ocean
  waves}}}}\ (\bibinfo  {publisher} {Academic Press},\ \bibinfo {year}
  {2010})\BibitemShut {NoStop}%
\bibitem [{\citenamefont {Chabchoub}\ \emph {et~al.}(2015)\citenamefont
  {Chabchoub}, \citenamefont {Kibler}, \citenamefont {Finot}, \citenamefont
  {Millot}, \citenamefont {Onorato}, \citenamefont {Dudley},\ and\
  \citenamefont {Babanin}}]{Chabchoub:15}%
  \BibitemOpen
  \bibfield  {author} {\bibinfo {author} {\bibfnamefont {A.}~\bibnamefont
  {Chabchoub}}, \bibinfo {author} {\bibfnamefont {B.}~\bibnamefont {Kibler}},
  \bibinfo {author} {\bibfnamefont {C.}~\bibnamefont {Finot}}, \bibinfo
  {author} {\bibfnamefont {G.}~\bibnamefont {Millot}}, \bibinfo {author}
  {\bibfnamefont {M.}~\bibnamefont {Onorato}}, \bibinfo {author} {\bibfnamefont
  {J.}~\bibnamefont {Dudley}}, \ and\ \bibinfo {author} {\bibfnamefont
  {A.}~\bibnamefont {Babanin}},\ }\href {\doibase
  http://dx.doi.org/10.1016/j.aop.2015.07.003} {\bibfield  {journal} {\bibinfo
  {journal} {Annals of Physics}\ }\textbf {\bibinfo {volume} {361}},\ \bibinfo
  {pages} {490 } (\bibinfo {year} {2015})}\BibitemShut {NoStop}%
\bibitem [{\citenamefont {Chabchoub}\ and\ \citenamefont
  {Grimshaw}(2016)}]{Chabchoub:16}%
  \BibitemOpen
  \bibfield  {author} {\bibinfo {author} {\bibfnamefont {A.}~\bibnamefont
  {Chabchoub}}\ and\ \bibinfo {author} {\bibfnamefont {R.~H.~J.}\ \bibnamefont
  {Grimshaw}},\ }\href {\doibase 10.3390/fluids1030023} {\bibfield  {journal}
  {\bibinfo  {journal} {Fluids}\ }\textbf {\bibinfo {volume} {1}} (\bibinfo
  {year} {2016}),\ 10.3390/fluids1030023}\BibitemShut {NoStop}%
\bibitem [{\citenamefont {Tracy}\ \emph {et~al.}(1984)\citenamefont {Tracy},
  \citenamefont {Chen},\ and\ \citenamefont {Lee}}]{Tracy:84}%
  \BibitemOpen
  \bibfield  {author} {\bibinfo {author} {\bibfnamefont {E.~R.}\ \bibnamefont
  {Tracy}}, \bibinfo {author} {\bibfnamefont {H.~H.}\ \bibnamefont {Chen}}, \
  and\ \bibinfo {author} {\bibfnamefont {Y.~C.}\ \bibnamefont {Lee}},\ }\href
  {\doibase 10.1103/PhysRevLett.53.218} {\bibfield  {journal} {\bibinfo
  {journal} {Phys. Rev. Lett.}\ }\textbf {\bibinfo {volume} {53}},\ \bibinfo
  {pages} {218} (\bibinfo {year} {1984})}\BibitemShut {NoStop}%
\bibitem [{\citenamefont {Grinevich}\ and\ \citenamefont
  {Santini}(2017)}]{Grinevich:17}%
  \BibitemOpen
  \bibfield  {author} {\bibinfo {author} {\bibfnamefont {P.~G.}\ \bibnamefont
  {Grinevich}}\ and\ \bibinfo {author} {\bibfnamefont {P.}~\bibnamefont
  {Santini}},\ }\href@noop {} {\bibfield  {journal} {\bibinfo  {journal} {Arxiv
  preprint arXiv:1707.05659}\ } (\bibinfo {year} {2017})}\BibitemShut {NoStop}%
\bibitem [{\citenamefont {Eastham}(1973)}]{Easthambook}%
  \BibitemOpen
  \bibfield  {author} {\bibinfo {author} {\bibfnamefont {M.~S.~P.}\
  \bibnamefont {Eastham}},\ }\href
  {http://gen.lib.rus.ec/book/index.php?md5=872D6D3BC656C9CCAC3104119A8B03E6}
  {\emph {\bibinfo {title} {The Spectral Theory of Periodic Differential
  Equations}}}\ (\bibinfo  {publisher} {Scottish Academic Press Ltd},\ \bibinfo
  {year} {1973})\BibitemShut {NoStop}%
\bibitem [{\citenamefont {Ablowitz}\ \emph {et~al.}(2001)\citenamefont
  {Ablowitz}, \citenamefont {Hammack}, \citenamefont {Henderson},\ and\
  \citenamefont {Schober}}]{Ablowitz:01}%
  \BibitemOpen
  \bibfield  {author} {\bibinfo {author} {\bibfnamefont {M.}~\bibnamefont
  {Ablowitz}}, \bibinfo {author} {\bibfnamefont {J.}~\bibnamefont {Hammack}},
  \bibinfo {author} {\bibfnamefont {D.}~\bibnamefont {Henderson}}, \ and\
  \bibinfo {author} {\bibfnamefont {C.}~\bibnamefont {Schober}},\ }\href@noop
  {} {\bibfield  {journal} {\bibinfo  {journal} {Physica D: Nonlinear
  Phenomena}\ }\textbf {\bibinfo {volume} {152}},\ \bibinfo {pages} {416}
  (\bibinfo {year} {2001})}\BibitemShut {NoStop}%
\bibitem [{\citenamefont {Islas}\ and\ \citenamefont
  {Schober}(2005)}]{Islas:05}%
  \BibitemOpen
  \bibfield  {author} {\bibinfo {author} {\bibfnamefont {A.~L.}\ \bibnamefont
  {Islas}}\ and\ \bibinfo {author} {\bibfnamefont {C.~M.}\ \bibnamefont
  {Schober}},\ }\href@noop {} {\bibfield  {journal} {\bibinfo  {journal}
  {Physics of Fluids}\ }\textbf {\bibinfo {volume} {17}},\ \bibinfo {eid}
  {031701} (\bibinfo {year} {2005})}\BibitemShut {NoStop}%
\bibitem [{\citenamefont {Calini}\ and\ \citenamefont
  {Schober}(2012)}]{Calini:12}%
  \BibitemOpen
  \bibfield  {author} {\bibinfo {author} {\bibfnamefont {A.}~\bibnamefont
  {Calini}}\ and\ \bibinfo {author} {\bibfnamefont {C.~M.}\ \bibnamefont
  {Schober}},\ }\href@noop {} {\bibfield  {journal} {\bibinfo  {journal}
  {Nonlinearity}\ }\textbf {\bibinfo {volume} {25}},\ \bibinfo {pages} {R99}
  (\bibinfo {year} {2012})}\BibitemShut {NoStop}%
\bibitem [{\citenamefont {El}\ \emph {et~al.}(2016)\citenamefont {El},
  \citenamefont {Khamis},\ and\ \citenamefont {Tovbis}}]{El:16}%
  \BibitemOpen
  \bibfield  {author} {\bibinfo {author} {\bibfnamefont {G.~A.}\ \bibnamefont
  {El}}, \bibinfo {author} {\bibfnamefont {E.~G.}\ \bibnamefont {Khamis}}, \
  and\ \bibinfo {author} {\bibfnamefont {A.}~\bibnamefont {Tovbis}},\ }\href
  {http://stacks.iop.org/0951-7715/29/i=9/a=2798} {\bibfield  {journal}
  {\bibinfo  {journal} {Nonlinearity}\ }\textbf {\bibinfo {volume} {29}},\
  \bibinfo {pages} {2798} (\bibinfo {year} {2016})}\BibitemShut {NoStop}%
\bibitem [{\citenamefont {Ma}\ and\ \citenamefont
  {Ablowitz}(1981)}]{ma1981periodic}%
  \BibitemOpen
  \bibfield  {author} {\bibinfo {author} {\bibfnamefont {Y.-C.}\ \bibnamefont
  {Ma}}\ and\ \bibinfo {author} {\bibfnamefont {M.~J.}\ \bibnamefont
  {Ablowitz}},\ }\href@noop {} {\bibfield  {journal} {\bibinfo  {journal}
  {Stud. Appl. Math.}\ }\textbf {\bibinfo {volume} {65}},\ \bibinfo {pages}
  {113} (\bibinfo {year} {1981})}\BibitemShut {NoStop}%
\bibitem [{\citenamefont {Randoux}\ \emph
  {et~al.}(2016{\natexlab{b}})\citenamefont {Randoux}, \citenamefont {Suret},\
  and\ \citenamefont {El}}]{Randoux:16b}%
  \BibitemOpen
  \bibfield  {author} {\bibinfo {author} {\bibfnamefont {S.}~\bibnamefont
  {Randoux}}, \bibinfo {author} {\bibfnamefont {P.}~\bibnamefont {Suret}}, \
  and\ \bibinfo {author} {\bibfnamefont {G.}~\bibnamefont {El}},\ }\href@noop
  {} {\bibfield  {journal} {\bibinfo  {journal} {Scientific reports}\ }\textbf
  {\bibinfo {volume} {6}},\ \bibinfo {pages} {29238} (\bibinfo {year}
  {2016}{\natexlab{b}})}\BibitemShut {NoStop}%
\bibitem [{\citenamefont {Biondini}\ and\ \citenamefont
  {Kovačič}(2014)}]{Biondini:14}%
  \BibitemOpen
  \bibfield  {author} {\bibinfo {author} {\bibfnamefont {G.}~\bibnamefont
  {Biondini}}\ and\ \bibinfo {author} {\bibfnamefont {G.}~\bibnamefont
  {Kovačič}},\ }\href {\doibase http://dx.doi.org/10.1063/1.4868483}
  {\bibfield  {journal} {\bibinfo  {journal} {J. Math. Phys.}\ }\textbf
  {\bibinfo {volume} {55}},\ \bibinfo {eid} {031506} (\bibinfo {year}
  {2014})}\BibitemShut {NoStop}%
\bibitem [{\citenamefont {Biondini}\ and\ \citenamefont
  {Fagerstrom}(2015)}]{Biondini:15}%
  \BibitemOpen
  \bibfield  {author} {\bibinfo {author} {\bibfnamefont {G.}~\bibnamefont
  {Biondini}}\ and\ \bibinfo {author} {\bibfnamefont {E.}~\bibnamefont
  {Fagerstrom}},\ }\href {\doibase http://dx.doi.org/10.1137/140965089}
  {\bibfield  {journal} {\bibinfo  {journal} {SIAM J. Appl. Math.}\ }\textbf
  {\bibinfo {volume} {75}},\ \bibinfo {pages} {136} (\bibinfo {year}
  {2015})}\BibitemShut {NoStop}%
\bibitem [{\citenamefont {Gelash}\ and\ \citenamefont
  {Zakharov}(2014)}]{Gelash:14}%
  \BibitemOpen
  \bibfield  {author} {\bibinfo {author} {\bibfnamefont {A.~A.}\ \bibnamefont
  {Gelash}}\ and\ \bibinfo {author} {\bibfnamefont {V.~E.}\ \bibnamefont
  {Zakharov}},\ }\href {http://stacks.iop.org/0951-7715/27/i=4/a=R1} {\bibfield
   {journal} {\bibinfo  {journal} {Nonlinearity}\ }\textbf {\bibinfo {volume}
  {27}},\ \bibinfo {pages} {R1} (\bibinfo {year} {2014})}\BibitemShut {NoStop}%
\bibitem [{\citenamefont {Boffetta}\ and\ \citenamefont
  {Osborne}(1992)}]{Boffetta:92}%
  \BibitemOpen
  \bibfield  {author} {\bibinfo {author} {\bibfnamefont {G.}~\bibnamefont
  {Boffetta}}\ and\ \bibinfo {author} {\bibfnamefont {A.}~\bibnamefont
  {Osborne}},\ }\href@noop {} {\bibfield  {journal} {\bibinfo  {journal} {J.
  Comp. Phys.}\ }\textbf {\bibinfo {volume} {102}},\ \bibinfo {pages} {252}
  (\bibinfo {year} {1992})}\BibitemShut {NoStop}%
\bibitem [{\citenamefont {Yousefi}\ and\ \citenamefont
  {Kschischang}(2014)}]{Yousefi:14}%
  \BibitemOpen
  \bibfield  {author} {\bibinfo {author} {\bibfnamefont {M.~I.}\ \bibnamefont
  {Yousefi}}\ and\ \bibinfo {author} {\bibfnamefont {F.~R.}\ \bibnamefont
  {Kschischang}},\ }\href@noop {} {\bibfield  {journal} {\bibinfo  {journal}
  {IEEE Transactions on Information Theory}\ }\textbf {\bibinfo {volume}
  {60}},\ \bibinfo {pages} {4329} (\bibinfo {year} {2014})}\BibitemShut
  {NoStop}%
\bibitem [{\citenamefont {Wahls}\ and\ \citenamefont {Poor}(2015)}]{Wahls:15}%
  \BibitemOpen
  \bibfield  {author} {\bibinfo {author} {\bibfnamefont {S.}~\bibnamefont
  {Wahls}}\ and\ \bibinfo {author} {\bibfnamefont {H.~V.}\ \bibnamefont
  {Poor}},\ }\href {\doibase 10.1109/TIT.2015.2485944} {\bibfield  {journal}
  {\bibinfo  {journal} {IEEE Transactions on Information Theory}\ }\textbf
  {\bibinfo {volume} {61}},\ \bibinfo {pages} {6957} (\bibinfo {year}
  {2015})}\BibitemShut {NoStop}%
\bibitem [{\citenamefont {Frumin}\ \emph {et~al.}(2015)\citenamefont {Frumin},
  \citenamefont {Belai}, \citenamefont {Podivilov},\ and\ \citenamefont
  {Shapiro}}]{Frumin:15}%
  \BibitemOpen
  \bibfield  {author} {\bibinfo {author} {\bibfnamefont {L.~L.}\ \bibnamefont
  {Frumin}}, \bibinfo {author} {\bibfnamefont {O.~V.}\ \bibnamefont {Belai}},
  \bibinfo {author} {\bibfnamefont {E.~V.}\ \bibnamefont {Podivilov}}, \ and\
  \bibinfo {author} {\bibfnamefont {D.~A.}\ \bibnamefont {Shapiro}},\ }\href
  {\doibase 10.1364/JOSAB.32.000290} {\bibfield  {journal} {\bibinfo  {journal}
  {J. Opt. Soc. Am. B}\ }\textbf {\bibinfo {volume} {32}},\ \bibinfo {pages}
  {290} (\bibinfo {year} {2015})}\BibitemShut {NoStop}%
\bibitem [{\citenamefont {Kamalian}\ \emph {et~al.}(2016)\citenamefont
  {Kamalian}, \citenamefont {Prilepsky}, \citenamefont {Le},\ and\
  \citenamefont {Turitsyn}}]{Kamalian:16}%
  \BibitemOpen
  \bibfield  {author} {\bibinfo {author} {\bibfnamefont {M.}~\bibnamefont
  {Kamalian}}, \bibinfo {author} {\bibfnamefont {J.~E.}\ \bibnamefont
  {Prilepsky}}, \bibinfo {author} {\bibfnamefont {S.~T.}\ \bibnamefont {Le}}, \
  and\ \bibinfo {author} {\bibfnamefont {S.~K.}\ \bibnamefont {Turitsyn}},\
  }\href@noop {} {\bibfield  {journal} {\bibinfo  {journal} {Optics express}\
  }\textbf {\bibinfo {volume} {24}},\ \bibinfo {pages} {18353} (\bibinfo {year}
  {2016})}\BibitemShut {NoStop}%
\bibitem [{\citenamefont {Turitsyn}\ \emph {et~al.}(2017)\citenamefont
  {Turitsyn}, \citenamefont {Prilepsky}, \citenamefont {Le}, \citenamefont
  {Wahls}, \citenamefont {Frumin}, \citenamefont {Kamalian},\ and\
  \citenamefont {Derevyanko}}]{Turitsyn:17}%
  \BibitemOpen
  \bibfield  {author} {\bibinfo {author} {\bibfnamefont {S.~K.}\ \bibnamefont
  {Turitsyn}}, \bibinfo {author} {\bibfnamefont {J.~E.}\ \bibnamefont
  {Prilepsky}}, \bibinfo {author} {\bibfnamefont {S.~T.}\ \bibnamefont {Le}},
  \bibinfo {author} {\bibfnamefont {S.}~\bibnamefont {Wahls}}, \bibinfo
  {author} {\bibfnamefont {L.~L.}\ \bibnamefont {Frumin}}, \bibinfo {author}
  {\bibfnamefont {M.}~\bibnamefont {Kamalian}}, \ and\ \bibinfo {author}
  {\bibfnamefont {S.~A.}\ \bibnamefont {Derevyanko}},\ }\href@noop {}
  {\bibfield  {journal} {\bibinfo  {journal} {Optica}\ }\textbf {\bibinfo
  {volume} {4}},\ \bibinfo {pages} {307} (\bibinfo {year} {2017})}\BibitemShut
  {NoStop}%
\bibitem [{\citenamefont {Slunyaev}(2006)}]{Slunyaev:06}%
  \BibitemOpen
  \bibfield  {author} {\bibinfo {author} {\bibfnamefont {A.}~\bibnamefont
  {Slunyaev}},\ }\href@noop {} {\bibfield  {journal} {\bibinfo  {journal}
  {European Journal of Mechanics-B/Fluids}\ }\textbf {\bibinfo {volume} {25}},\
  \bibinfo {pages} {621} (\bibinfo {year} {2006})}\BibitemShut {NoStop}%
\bibitem [{\citenamefont {Osborne}(1993)}]{Osborne:93}%
  \BibitemOpen
  \bibfield  {author} {\bibinfo {author} {\bibfnamefont {A.}~\bibnamefont
  {Osborne}},\ }\href@noop {} {\bibfield  {journal} {\bibinfo  {journal}
  {Physical review letters}\ }\textbf {\bibinfo {volume} {71}},\ \bibinfo
  {pages} {3115} (\bibinfo {year} {1993})}\BibitemShut {NoStop}%
\bibitem [{\citenamefont {Osborne}(1995{\natexlab{a}})}]{Osborne:95}%
  \BibitemOpen
  \bibfield  {author} {\bibinfo {author} {\bibfnamefont {A.}~\bibnamefont
  {Osborne}},\ }\href@noop {} {\bibfield  {journal} {\bibinfo  {journal}
  {Physical Review E}\ }\textbf {\bibinfo {volume} {52}},\ \bibinfo {pages}
  {1105} (\bibinfo {year} {1995}{\natexlab{a}})}\BibitemShut {NoStop}%
\bibitem [{\citenamefont {Osborne}(1995{\natexlab{b}})}]{Osborne:95b}%
  \BibitemOpen
  \bibfield  {author} {\bibinfo {author} {\bibfnamefont {A.}~\bibnamefont
  {Osborne}},\ }\href {\doibase http://dx.doi.org/10.1016/0167-2789(95)00089-M}
  {\bibfield  {journal} {\bibinfo  {journal} {Physica D: Nonlinear Phenomena}\
  }\textbf {\bibinfo {volume} {86}},\ \bibinfo {pages} {81 } (\bibinfo {year}
  {1995}{\natexlab{b}})}\BibitemShut {NoStop}%
\bibitem [{\citenamefont {Costa}\ \emph {et~al.}(2014)\citenamefont {Costa},
  \citenamefont {Osborne}, \citenamefont {Resio}, \citenamefont {Alessio},
  \citenamefont {Chriv\`{\i}}, \citenamefont {Saggese}, \citenamefont
  {Bellomo},\ and\ \citenamefont {Long}}]{Osborne:14}%
  \BibitemOpen
  \bibfield  {author} {\bibinfo {author} {\bibfnamefont {A.}~\bibnamefont
  {Costa}}, \bibinfo {author} {\bibfnamefont {A.~R.}\ \bibnamefont {Osborne}},
  \bibinfo {author} {\bibfnamefont {D.~T.}\ \bibnamefont {Resio}}, \bibinfo
  {author} {\bibfnamefont {S.}~\bibnamefont {Alessio}}, \bibinfo {author}
  {\bibfnamefont {E.}~\bibnamefont {Chriv\`{\i}}}, \bibinfo {author}
  {\bibfnamefont {E.}~\bibnamefont {Saggese}}, \bibinfo {author} {\bibfnamefont
  {K.}~\bibnamefont {Bellomo}}, \ and\ \bibinfo {author} {\bibfnamefont
  {C.~E.}\ \bibnamefont {Long}},\ }\href {\doibase
  10.1103/PhysRevLett.113.108501} {\bibfield  {journal} {\bibinfo  {journal}
  {Phys. Rev. Lett.}\ }\textbf {\bibinfo {volume} {113}},\ \bibinfo {pages}
  {108501} (\bibinfo {year} {2014})}\BibitemShut {NoStop}%
\bibitem [{\citenamefont {Islas}\ and\ \citenamefont
  {Schober}(2011)}]{Islas:11}%
  \BibitemOpen
  \bibfield  {author} {\bibinfo {author} {\bibfnamefont {A.}~\bibnamefont
  {Islas}}\ and\ \bibinfo {author} {\bibfnamefont {C.}~\bibnamefont
  {Schober}},\ }\href@noop {} {\bibfield  {journal} {\bibinfo  {journal}
  {Physica D: Nonlinear Phenomena}\ }\textbf {\bibinfo {volume} {240}},\
  \bibinfo {pages} {1041} (\bibinfo {year} {2011})}\BibitemShut {NoStop}%
\bibitem [{\citenamefont {Prilepsky}\ \emph {et~al.}(2014)\citenamefont
  {Prilepsky}, \citenamefont {Derevyanko}, \citenamefont {Blow}, \citenamefont
  {Gabitov},\ and\ \citenamefont {Turitsyn}}]{Prilepsky:14}%
  \BibitemOpen
  \bibfield  {author} {\bibinfo {author} {\bibfnamefont {J.~E.}\ \bibnamefont
  {Prilepsky}}, \bibinfo {author} {\bibfnamefont {S.~A.}\ \bibnamefont
  {Derevyanko}}, \bibinfo {author} {\bibfnamefont {K.~J.}\ \bibnamefont
  {Blow}}, \bibinfo {author} {\bibfnamefont {I.}~\bibnamefont {Gabitov}}, \
  and\ \bibinfo {author} {\bibfnamefont {S.~K.}\ \bibnamefont {Turitsyn}},\
  }\href {\doibase 10.1103/PhysRevLett.113.013901} {\bibfield  {journal}
  {\bibinfo  {journal} {Phys. Rev. Lett.}\ }\textbf {\bibinfo {volume} {113}},\
  \bibinfo {pages} {013901} (\bibinfo {year} {2014})}\BibitemShut {NoStop}%
\bibitem [{\citenamefont {Frumin}\ \emph {et~al.}(2017)\citenamefont {Frumin},
  \citenamefont {Gelash},\ and\ \citenamefont {Turitsyn}}]{Frumin:17}%
  \BibitemOpen
  \bibfield  {author} {\bibinfo {author} {\bibfnamefont {L.}~\bibnamefont
  {Frumin}}, \bibinfo {author} {\bibfnamefont {A.}~\bibnamefont {Gelash}}, \
  and\ \bibinfo {author} {\bibfnamefont {S.}~\bibnamefont {Turitsyn}},\
  }\href@noop {} {\bibfield  {journal} {\bibinfo  {journal} {Physical Review
  Letters}\ }\textbf {\bibinfo {volume} {118}},\ \bibinfo {pages} {223901}
  (\bibinfo {year} {2017})}\BibitemShut {NoStop}%
\bibitem [{\citenamefont {B\"ohm}\ and\ \citenamefont
  {Mitschke}(2006)}]{Bohm:06}%
  \BibitemOpen
  \bibfield  {author} {\bibinfo {author} {\bibfnamefont {M.}~\bibnamefont
  {B\"ohm}}\ and\ \bibinfo {author} {\bibfnamefont {F.}~\bibnamefont
  {Mitschke}},\ }\href {\doibase 10.1103/PhysRevE.73.066615} {\bibfield
  {journal} {\bibinfo  {journal} {Phys. Rev. E}\ }\textbf {\bibinfo {volume}
  {73}},\ \bibinfo {pages} {066615} (\bibinfo {year} {2006})}\BibitemShut
  {NoStop}%
\bibitem [{\citenamefont {B{\"o}hm}\ and\ \citenamefont
  {Mitschke}(2007)}]{Bohm:07}%
  \BibitemOpen
  \bibfield  {author} {\bibinfo {author} {\bibfnamefont {M.}~\bibnamefont
  {B{\"o}hm}}\ and\ \bibinfo {author} {\bibfnamefont {F.}~\bibnamefont
  {Mitschke}},\ }\href {\doibase 10.1007/s00340-006-2513-6} {\bibfield
  {journal} {\bibinfo  {journal} {Applied Physics B}\ }\textbf {\bibinfo
  {volume} {86}},\ \bibinfo {pages} {407} (\bibinfo {year} {2007})}\BibitemShut
  {NoStop}%
\bibitem [{\citenamefont {Mitschke}\ \emph {et~al.}(2017)\citenamefont
  {Mitschke}, \citenamefont {Mahnke},\ and\ \citenamefont
  {Hause}}]{Mitschke:17}%
  \BibitemOpen
  \bibfield  {author} {\bibinfo {author} {\bibfnamefont {F.}~\bibnamefont
  {Mitschke}}, \bibinfo {author} {\bibfnamefont {C.}~\bibnamefont {Mahnke}}, \
  and\ \bibinfo {author} {\bibfnamefont {A.}~\bibnamefont {Hause}},\ }\href
  {\doibase 10.3390/app7060635} {\bibfield  {journal} {\bibinfo  {journal}
  {Applied Sciences}\ }\textbf {\bibinfo {volume} {7}},\ \bibinfo {pages} {635}
  (\bibinfo {year} {2017})}\BibitemShut {NoStop}%
\bibitem [{\citenamefont {Frisquet}\ \emph {et~al.}(2014)\citenamefont
  {Frisquet}, \citenamefont {Chabchoub}, \citenamefont {Fatome}, \citenamefont
  {Finot}, \citenamefont {Kibler},\ and\ \citenamefont {Millot}}]{Frisquet:14}%
  \BibitemOpen
  \bibfield  {author} {\bibinfo {author} {\bibfnamefont {B.}~\bibnamefont
  {Frisquet}}, \bibinfo {author} {\bibfnamefont {A.}~\bibnamefont {Chabchoub}},
  \bibinfo {author} {\bibfnamefont {J.}~\bibnamefont {Fatome}}, \bibinfo
  {author} {\bibfnamefont {C.}~\bibnamefont {Finot}}, \bibinfo {author}
  {\bibfnamefont {B.}~\bibnamefont {Kibler}}, \ and\ \bibinfo {author}
  {\bibfnamefont {G.}~\bibnamefont {Millot}},\ }\href {\doibase
  10.1103/PhysRevA.89.023821} {\bibfield  {journal} {\bibinfo  {journal} {Phys.
  Rev. A}\ }\textbf {\bibinfo {volume} {89}},\ \bibinfo {pages} {023821}
  (\bibinfo {year} {2014})}\BibitemShut {NoStop}%
\bibitem [{\citenamefont {Dudley}\ \emph {et~al.}(2009)\citenamefont {Dudley},
  \citenamefont {Genty}, \citenamefont {Dias}, \citenamefont {Kibler},\ and\
  \citenamefont {Akhmediev}}]{Dudley:09}%
  \BibitemOpen
  \bibfield  {author} {\bibinfo {author} {\bibfnamefont {J.~M.}\ \bibnamefont
  {Dudley}}, \bibinfo {author} {\bibfnamefont {G.}~\bibnamefont {Genty}},
  \bibinfo {author} {\bibfnamefont {F.}~\bibnamefont {Dias}}, \bibinfo {author}
  {\bibfnamefont {B.}~\bibnamefont {Kibler}}, \ and\ \bibinfo {author}
  {\bibfnamefont {N.}~\bibnamefont {Akhmediev}},\ }\href {\doibase
  10.1364/OE.17.021497} {\bibfield  {journal} {\bibinfo  {journal} {Opt.
  Express}\ }\textbf {\bibinfo {volume} {17}},\ \bibinfo {pages} {21497}
  (\bibinfo {year} {2009})}\BibitemShut {NoStop}%
\bibitem [{\citenamefont {Mussot}\ \emph {et~al.}(2018)\citenamefont {Mussot},
  \citenamefont {Naveau}, \citenamefont {Conforti}, \citenamefont {Kudlinski},
  \citenamefont {Copie}, \citenamefont {Szriftgiser},\ and\ \citenamefont
  {Trillo}}]{Mussot:18}%
  \BibitemOpen
  \bibfield  {author} {\bibinfo {author} {\bibfnamefont {A.}~\bibnamefont
  {Mussot}}, \bibinfo {author} {\bibfnamefont {C.}~\bibnamefont {Naveau}},
  \bibinfo {author} {\bibfnamefont {M.}~\bibnamefont {Conforti}}, \bibinfo
  {author} {\bibfnamefont {A.}~\bibnamefont {Kudlinski}}, \bibinfo {author}
  {\bibfnamefont {F.}~\bibnamefont {Copie}}, \bibinfo {author} {\bibfnamefont
  {P.}~\bibnamefont {Szriftgiser}}, \ and\ \bibinfo {author} {\bibfnamefont
  {S.}~\bibnamefont {Trillo}},\ }\href {\doibase 10.1038/s41566-018-0136-1}
  {\bibfield  {journal} {\bibinfo  {journal} {Nature Photonics}\ }\textbf
  {\bibinfo {volume} {12}},\ \bibinfo {pages} {303} (\bibinfo {year}
  {2018})}\BibitemShut {NoStop}%
\bibitem [{\citenamefont {Ablowitz}\ \emph {et~al.}(1996)\citenamefont
  {Ablowitz}, \citenamefont {Herbst},\ and\ \citenamefont
  {Schober}}]{Ablowitz:96}%
  \BibitemOpen
  \bibfield  {author} {\bibinfo {author} {\bibfnamefont {M.}~\bibnamefont
  {Ablowitz}}, \bibinfo {author} {\bibfnamefont {B.}~\bibnamefont {Herbst}}, \
  and\ \bibinfo {author} {\bibfnamefont {C.}~\bibnamefont {Schober}},\ }\href
  {\doibase https://doi.org/10.1016/0378-4371(95)00434-3} {\bibfield  {journal}
  {\bibinfo  {journal} {Physica A: Statistical Mechanics and its Applications}\
  }\textbf {\bibinfo {volume} {228}},\ \bibinfo {pages} {212 } (\bibinfo {year}
  {1996})}\BibitemShut {NoStop}%
\bibitem [{\citenamefont {Akhmediev}\ \emph
  {et~al.}(2009{\natexlab{d}})\citenamefont {Akhmediev}, \citenamefont
  {Soto-Crespo},\ and\ \citenamefont {Ankiewicz}}]{Akhmediev:09d}%
  \BibitemOpen
  \bibfield  {author} {\bibinfo {author} {\bibfnamefont {N.}~\bibnamefont
  {Akhmediev}}, \bibinfo {author} {\bibfnamefont {J.~M.}\ \bibnamefont
  {Soto-Crespo}}, \ and\ \bibinfo {author} {\bibfnamefont {A.}~\bibnamefont
  {Ankiewicz}},\ }\href {\doibase 10.1103/PhysRevA.80.043818} {\bibfield
  {journal} {\bibinfo  {journal} {Phys. Rev. A}\ }\textbf {\bibinfo {volume}
  {80}},\ \bibinfo {pages} {043818} (\bibinfo {year}
  {2009}{\natexlab{d}})}\BibitemShut {NoStop}%
\bibitem [{\citenamefont {Kibler}(2017)}]{shaping_light_book_kibler}%
  \BibitemOpen
  \bibfield  {author} {\bibinfo {author} {\bibfnamefont {B.}~\bibnamefont
  {Kibler}},\ }\href@noop {} {\emph {\bibinfo {title} {Shaping Light in
  Nonlinear Optical Fibers-Chapter 10: Rogue breather structures in nonlinear
  systems with an emphasis on optical fibers as testbeds}}}\ (\bibinfo
  {publisher} {John Wiley and Sons, Inc},\ \bibinfo {year} {2017})\BibitemShut
  {NoStop}%
\bibitem [{\citenamefont {Bertola}\ \emph {et~al.}(2016)\citenamefont
  {Bertola}, \citenamefont {El},\ and\ \citenamefont {Tovbis}}]{Bertola:16}%
  \BibitemOpen
  \bibfield  {author} {\bibinfo {author} {\bibfnamefont {M.}~\bibnamefont
  {Bertola}}, \bibinfo {author} {\bibfnamefont {G.~A.}\ \bibnamefont {El}}, \
  and\ \bibinfo {author} {\bibfnamefont {A.}~\bibnamefont {Tovbis}},\ }\href
  {http://rspa.royalsocietypublishing.org/content/472/2194/20160340} {\bibfield
   {journal} {\bibinfo  {journal} {Proceedings of the Royal Society of London
  A: Mathematical, Physical and Engineering Sciences}\ }\textbf {\bibinfo
  {volume} {472}} (\bibinfo {year} {2016})}\BibitemShut {NoStop}%
\bibitem [{\citenamefont {Hammani}\ \emph
  {et~al.}(2011{\natexlab{b}})\citenamefont {Hammani}, \citenamefont {Kibler},
  \citenamefont {Finot}, \citenamefont {Morin}, \citenamefont {Fatome},
  \citenamefont {Dudley},\ and\ \citenamefont {Millot}}]{Hammani:11b}%
  \BibitemOpen
  \bibfield  {author} {\bibinfo {author} {\bibfnamefont {K.}~\bibnamefont
  {Hammani}}, \bibinfo {author} {\bibfnamefont {B.}~\bibnamefont {Kibler}},
  \bibinfo {author} {\bibfnamefont {C.}~\bibnamefont {Finot}}, \bibinfo
  {author} {\bibfnamefont {P.}~\bibnamefont {Morin}}, \bibinfo {author}
  {\bibfnamefont {J.}~\bibnamefont {Fatome}}, \bibinfo {author} {\bibfnamefont
  {J.~M.}\ \bibnamefont {Dudley}}, \ and\ \bibinfo {author} {\bibfnamefont
  {G.}~\bibnamefont {Millot}},\ }\href {\doibase 10.1364/OL.36.000112}
  {\bibfield  {journal} {\bibinfo  {journal} {Opt. Lett.}\ }\textbf {\bibinfo
  {volume} {36}},\ \bibinfo {pages} {112} (\bibinfo {year}
  {2011}{\natexlab{b}})}\BibitemShut {NoStop}%
\bibitem [{\citenamefont {Erkintalo}\ \emph {et~al.}(2011)\citenamefont
  {Erkintalo}, \citenamefont {Hammani}, \citenamefont {Kibler}, \citenamefont
  {Finot}, \citenamefont {Akhmediev}, \citenamefont {Dudley},\ and\
  \citenamefont {Genty}}]{Erkintalo:11}%
  \BibitemOpen
  \bibfield  {author} {\bibinfo {author} {\bibfnamefont {M.}~\bibnamefont
  {Erkintalo}}, \bibinfo {author} {\bibfnamefont {K.}~\bibnamefont {Hammani}},
  \bibinfo {author} {\bibfnamefont {B.}~\bibnamefont {Kibler}}, \bibinfo
  {author} {\bibfnamefont {C.}~\bibnamefont {Finot}}, \bibinfo {author}
  {\bibfnamefont {N.}~\bibnamefont {Akhmediev}}, \bibinfo {author}
  {\bibfnamefont {J.~M.}\ \bibnamefont {Dudley}}, \ and\ \bibinfo {author}
  {\bibfnamefont {G.}~\bibnamefont {Genty}},\ }\href {\doibase
  10.1103/PhysRevLett.107.253901} {\bibfield  {journal} {\bibinfo  {journal}
  {Phys. Rev. Lett.}\ }\textbf {\bibinfo {volume} {107}},\ \bibinfo {pages}
  {253901} (\bibinfo {year} {2011})}\BibitemShut {NoStop}%
\bibitem [{\citenamefont {Kimmoun}\ \emph {et~al.}(2017)\citenamefont
  {Kimmoun}, \citenamefont {Hsu}, \citenamefont {Kibler},\ and\ \citenamefont
  {Chabchoub}}]{Kimmoun:17}%
  \BibitemOpen
  \bibfield  {author} {\bibinfo {author} {\bibfnamefont {O.}~\bibnamefont
  {Kimmoun}}, \bibinfo {author} {\bibfnamefont {H.~C.}\ \bibnamefont {Hsu}},
  \bibinfo {author} {\bibfnamefont {B.}~\bibnamefont {Kibler}}, \ and\ \bibinfo
  {author} {\bibfnamefont {A.}~\bibnamefont {Chabchoub}},\ }\href {\doibase
  10.1103/PhysRevE.96.022219} {\bibfield  {journal} {\bibinfo  {journal} {Phys.
  Rev. E}\ }\textbf {\bibinfo {volume} {96}},\ \bibinfo {pages} {022219}
  (\bibinfo {year} {2017})}\BibitemShut {NoStop}%
\bibitem [{\citenamefont {Kimmoun}\ \emph {et~al.}(2016)\citenamefont
  {Kimmoun}, \citenamefont {Hsu}, \citenamefont {Branger}, \citenamefont {Li},
  \citenamefont {Chen}, \citenamefont {Kharif},\ and\ \citenamefont
  {Chabchoub}}]{Kimmoun:16}%
  \BibitemOpen
  \bibfield  {author} {\bibinfo {author} {\bibfnamefont {O.}~\bibnamefont
  {Kimmoun}}, \bibinfo {author} {\bibfnamefont {H.~C.}\ \bibnamefont {Hsu}},
  \bibinfo {author} {\bibfnamefont {H.}~\bibnamefont {Branger}}, \bibinfo
  {author} {\bibfnamefont {M.~S.}\ \bibnamefont {Li}}, \bibinfo {author}
  {\bibfnamefont {Y.~Y.}\ \bibnamefont {Chen}}, \bibinfo {author}
  {\bibfnamefont {C.}~\bibnamefont {Kharif}}, \ and\ \bibinfo {author}
  {\bibfnamefont {A.}~\bibnamefont {Chabchoub}},\ }\href@noop {} {\bibfield
  {journal} {\bibinfo  {journal} {Scientific reports}\ }\textbf {\bibinfo
  {volume} {6}},\ \bibinfo {pages} {28516} (\bibinfo {year}
  {2016})}\BibitemShut {NoStop}%
\bibitem [{\citenamefont {Shemer}\ \emph {et~al.}(2002)\citenamefont {Shemer},
  \citenamefont {Kit},\ and\ \citenamefont {Jiao}}]{Shemer:02}%
  \BibitemOpen
  \bibfield  {author} {\bibinfo {author} {\bibfnamefont {L.}~\bibnamefont
  {Shemer}}, \bibinfo {author} {\bibfnamefont {E.}~\bibnamefont {Kit}}, \ and\
  \bibinfo {author} {\bibfnamefont {H.}~\bibnamefont {Jiao}},\ }\href@noop {}
  {\bibfield  {journal} {\bibinfo  {journal} {Physics of fluids}\ }\textbf
  {\bibinfo {volume} {14}},\ \bibinfo {pages} {3380} (\bibinfo {year}
  {2002})}\BibitemShut {NoStop}%
\bibitem [{\citenamefont {Chabchoub}\ \emph
  {et~al.}(2012{\natexlab{c}})\citenamefont {Chabchoub}, \citenamefont
  {Akhmediev},\ and\ \citenamefont {Hoffmann}}]{Chabchoub:12c}%
  \BibitemOpen
  \bibfield  {author} {\bibinfo {author} {\bibfnamefont {A.}~\bibnamefont
  {Chabchoub}}, \bibinfo {author} {\bibfnamefont {N.}~\bibnamefont
  {Akhmediev}}, \ and\ \bibinfo {author} {\bibfnamefont {N.~P.}\ \bibnamefont
  {Hoffmann}},\ }\href {\doibase 10.1103/PhysRevE.86.016311} {\bibfield
  {journal} {\bibinfo  {journal} {Phys. Rev. E}\ }\textbf {\bibinfo {volume}
  {86}},\ \bibinfo {pages} {016311} (\bibinfo {year}
  {2012}{\natexlab{c}})}\BibitemShut {NoStop}%
\bibitem [{\citenamefont {Akhmediev}\ \emph
  {et~al.}(2011{\natexlab{a}})\citenamefont {Akhmediev}, \citenamefont
  {Ankiewicz}, \citenamefont {Soto-Crespo},\ and\ \citenamefont
  {Dudley}}]{Akhmediev:11a}%
  \BibitemOpen
  \bibfield  {author} {\bibinfo {author} {\bibfnamefont {N.}~\bibnamefont
  {Akhmediev}}, \bibinfo {author} {\bibfnamefont {A.}~\bibnamefont
  {Ankiewicz}}, \bibinfo {author} {\bibfnamefont {J.}~\bibnamefont
  {Soto-Crespo}}, \ and\ \bibinfo {author} {\bibfnamefont {J.~M.}\ \bibnamefont
  {Dudley}},\ }\href {\doibase https://doi.org/10.1016/j.physleta.2010.11.044}
  {\bibfield  {journal} {\bibinfo  {journal} {Physics Letters A}\ }\textbf
  {\bibinfo {volume} {375}},\ \bibinfo {pages} {775 } (\bibinfo {year}
  {2011}{\natexlab{a}})}\BibitemShut {NoStop}%
\bibitem [{\citenamefont {Akhmediev}\ \emph
  {et~al.}(2011{\natexlab{b}})\citenamefont {Akhmediev}, \citenamefont
  {Ankiewicz}, \citenamefont {Soto-Crespo},\ and\ \citenamefont
  {Dudley}}]{Akhmediev:11b}%
  \BibitemOpen
  \bibfield  {author} {\bibinfo {author} {\bibfnamefont {N.}~\bibnamefont
  {Akhmediev}}, \bibinfo {author} {\bibfnamefont {A.}~\bibnamefont
  {Ankiewicz}}, \bibinfo {author} {\bibfnamefont {J.}~\bibnamefont
  {Soto-Crespo}}, \ and\ \bibinfo {author} {\bibfnamefont {J.~M.}\ \bibnamefont
  {Dudley}},\ }\href@noop {} {\bibfield  {journal} {\bibinfo  {journal}
  {Physics Letters A}\ }\textbf {\bibinfo {volume} {375}},\ \bibinfo {pages}
  {541} (\bibinfo {year} {2011}{\natexlab{b}})}\BibitemShut {NoStop}%
\bibitem [{\citenamefont {Cousins}\ and\ \citenamefont
  {Sapsis}(2016)}]{Cousins:16}%
  \BibitemOpen
  \bibfield  {author} {\bibinfo {author} {\bibfnamefont {W.}~\bibnamefont
  {Cousins}}\ and\ \bibinfo {author} {\bibfnamefont {T.~P.}\ \bibnamefont
  {Sapsis}},\ }\href@noop {} {\bibfield  {journal} {\bibinfo  {journal}
  {Journal of Fluid Mechanics}\ }\textbf {\bibinfo {volume} {790}},\ \bibinfo
  {pages} {368} (\bibinfo {year} {2016})}\BibitemShut {NoStop}%
\end{thebibliography}
%

\end{document}